\documentclass[manuscript, screen]{acmart}

\AtBeginDocument{%
  \providecommand\BibTeX{{%
    Bib\TeX}}}

\setcopyright{acmlicensed}
\copyrightyear{2018}
\acmYear{2018}
\acmDOI{XXXXXXX.XXXXXXX}

\acmConference[Conference acronym 'XX]{Make sure to enter the correct
  conference title from your rights confirmation emai}{June 03--05,
  2018}{Woodstock, NY}

\acmISBN{978-1-4503-XXXX-X/18/06}

\usepackage{multirow}
\usepackage{enumerate}

\newcommand{\red}[1]{\textcolor{red}{#1}}

\PassOptionsToPackage{numbers, sort&compress}{natbib}
\usepackage{paralist, tabularx}
\usepackage{amsmath}
\usepackage{float}
\usepackage{hyperref}
\hypersetup{colorlinks,urlcolor=blue,citecolor=blue,breaklinks=true}
\usepackage{url}

\usepackage{breakurl}
\usepackage{cellspace}
\usepackage{booktabs,adjustbox}
\usepackage{tcolorbox}
\newtcolorbox{mybox}{colback=black!5!white,colframe=black,bottomrule=.25mm,toprule=.25mm,leftrule=.25mm,rightrule=.25mm,left=.25mm,right=.25mm,top=-.25mm,bottom=-.25mm}
\usepackage{wrapfig}
\usepackage{amsfonts}
\usepackage{mathtools} 
\usepackage[english]{babel}
\usepackage{amsthm}
\usepackage{tabularx}
\usepackage{algorithm}
\usepackage{enumitem}
\usepackage{algpseudocode}

\usepackage{multirow}
\usepackage{array}
\usepackage{caption,subcaption}
\usepackage{listings}
\usepackage{diagbox}
\usepackage{tablefootnote}
\usepackage{lscape}

\def\BibTeX{{\rm B\kern-.05em{\sc i\kern-.025em b}\kern-.08em
    T\kern-.1667em\lower.7ex\hbox{E}\kern-.125emX}}

\makeatletter
\def\addlegendimage{\pgfplots@addlegendimage}
\makeatother

\usepackage{pifont}
\usepackage{comment}

\usepackage[utf8]{inputenc}
\newcommand{\paragraphb}[1]{\noindent{\bf #1} }

\usepackage{colortbl}

\definecolor{auburn}{rgb}{0.43, 0.21, 0.1}
\definecolor{burgundy}{rgb}{0.5, 0.0, 0.13}
\ifdefined\ToCheck
    \newcommand{\tocheck}[1]{\red{\bf@Amir}: \textcolor{cyan}{#1}}
\else
    \newcommand{\tocheck}[1]{{#1}}
\fi
\ifdefined\ToCheckAmin
    \newcommand{\tocheckAmir}[1]{\red{\bf@Amir}: \textcolor{cyan}{#1}}
\else
    \newcommand{\tocheckAmir}[1]{{}}
\fi

\makeatletter
\newcommand*\titleheader[1]{\gdef\@titleheader{#1}}
\AtBeginDocument{%
  \let\st@red@title\@title
  \def\@title{%
    \bgroup\small\large\centering\@titleheader\par\egroup
    \vskip.5em\st@red@title}
}
\makeatother
\usepackage{etoolbox}
\makeatletter
\patchcmd{\@maketitle}
  {\addvspace{0.5\baselineskip}\egroup}
  {\addvspace{-1em}\egroup}
  {}
  {}
\makeatother

\begin{document}

\title{Decoding FL Defenses: Systemization, Pitfalls, and Remedies}





\author{Momin Ahmad Khan}
\email{makhan@umass.edu}
\affiliation{%
  \institution{University of Massachusetts Amherst}
  \city{Amherst}
  \state{Massachusetts}
  \country{USA}
}

\author{Virat Shejwalkar}
\affiliation{%
  \institution{Google}
  \city{San Jose}
  \country{USA}}
\email{vshejwalkar@google.com}

\author{Yasra Chandio}
\email{ychandio@umass.edu}
\affiliation{%
  \institution{University of Massachusetts Amherst}
  \city{Amherst}
  \state{Massachusetts}
  \country{USA}
}

\author{Amir Houmansadr}
\email{amir@cs.umass.edu}
\affiliation{%
  \institution{University of Massachusetts Amherst}
  \city{Amherst}
  \state{Massachusetts}
  \country{USA}
}

\author{Fatima Muhammad Anwar}
\email{fanwar@umass.edu}
\affiliation{%
  \institution{University of Massachusetts Amherst}
  \city{Amherst}
  \state{Massachusetts}
  \country{USA}
}

\begin{abstract}\label{sec:abstract}
While the community has designed various defenses to counter the threat of poisoning attacks in Federated Learning (FL), there are no guidelines for evaluating these defenses. These defenses are prone to subtle pitfalls in their experimental setups that lead to a false sense of security, rendering them unsuitable for practical deployment.
In this paper, we systematically understand, identify, and provide a better approach to address these challenges.
First, we design a comprehensive systemization of FL defenses along three dimensions: 
i) how client updates are processed, ii) what the server knows, and iii) at what stage the defense is applied.
Next, we thoroughly survey 50 top-tier defense papers and identify the commonly used components in their evaluation setups. Based on this survey, we uncover six distinct pitfalls and study their prevalence. For example, we discover that around $30\%$ of these works solely use the intrinsically robust MNIST dataset, and $40\%$ employ simplistic attacks, which may inadvertently portray their defense as robust. 
Using three representative defenses as case studies, we perform a critical reevaluation to study the impact of the identified pitfalls and show how they lead to incorrect conclusions about robustness. 
We provide actionable recommendations to help researchers overcome each pitfall.
\end{abstract}

\maketitle

\section{Introduction}\label{sec:introduction}
Federated learning (FL)~\cite{mcmahan2017communication} is an emerging approach in machine learning (ML) where multiple data owners, called \emph{clients}, collaboratively train a shared model, known as the \emph{global model}, while keeping their individual training data private.
The central \emph{server} (service provider) iteratively aggregates \emph{model updates} from each client, which are generated based on their local data. The server merges these updates using an \emph{aggregation rule} (AGR) and uses them to update the global model. Following each training iteration (also known as \emph{round}), the refined global model is distributed to the clients participating in the next round. Prominent distributed platforms such as Google's Gboard~\cite{gboard} for next-word prediction, Apple's Siri~\cite{technologyreviewApplePersonalizes} for automatic speech recognition~\cite{paulik2021federated}, and WeBank~\cite{webankcredit} for credit risk predictions, have adopted this FL mechanism. Its intrinsic characteristic of promoting collaboration while preserving privacy has rendered it indispensable in critical applications, notably in medical diagnosis~\cite{feki2021federated, ku2022privacy, qayyum2022collaborative}, activity recognition~\cite{zhao2022multimodal, ek2020evaluation, ouyang2021clusterfl, sozinov2018human}, and next-character prediction~\cite{sun2022fedsea}.


\begin{figure}
\centering
\includegraphics[width=0.89\textwidth]{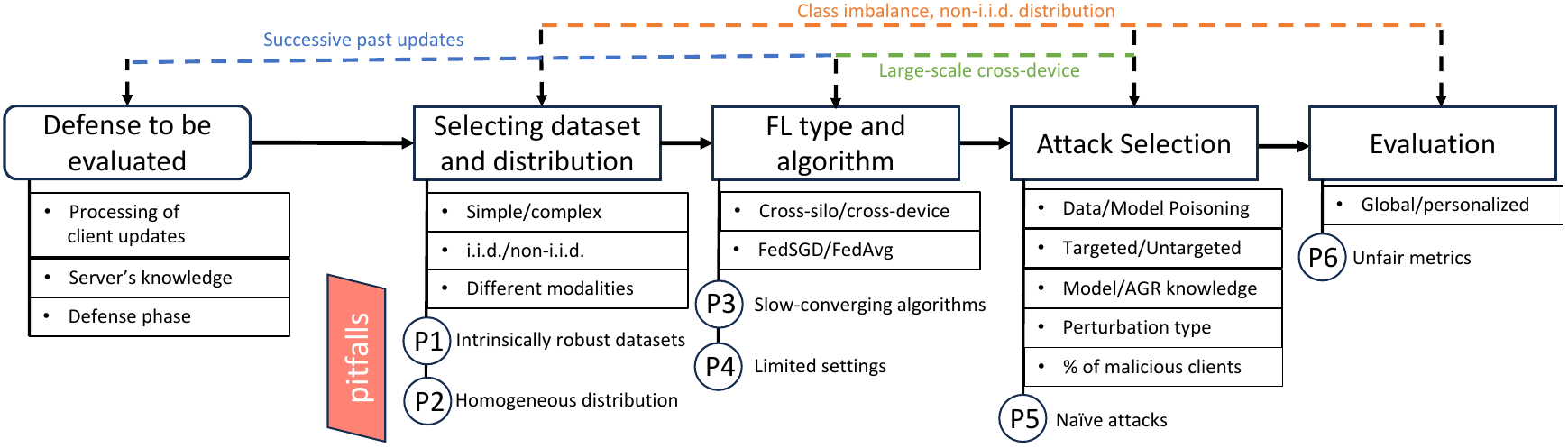}
\caption{
FL defense evaluation pipeline. We display common choices for each stage in the pipeline, e.g., FedSGD or FedAvg as the FL algorithm, and highlight the associated pitfall, i.e., using a slow-converging algorithm. We also indicate interdependencies between stages, e.g., large-scale and cross-device in FL type limit the number of malicious clients in the attack.
}
\label{fig:fl_flowchart}
\end{figure}

FL is gaining popularity due to its privacy-preserving and collaborative nature, yet it faces vulnerabilities to \emph{poisoning attacks}~\cite{fang2020local, shejwalkar2021manipulating, shejwalkar2022back, wang2020attack}, where malicious or \emph{compromised clients} intentionally corrupt FL training and \emph{poison} the global model. This can result in a poisoned model that performs poorly on all inputs in \emph{untargeted poisoning} attacks or on specific inputs in \emph{targeted poisoning} attacks. To address these threats, the community has developed various defense mechanisms. Robust AGRs such as Multi-krum~\cite{blanchard2017machine} and Trimmed-mean~\cite{yin2018byzantine}, detect and discard malicious updates. Certified defenses like CRFL~\cite{xie2021crfl} and Ensemble FL~\cite{cao2021provably} provide robustness certifications. Tools like FLDetector~\cite{zhang2022fldetector} proactively identify and remove malicious clients during training. Meanwhile, FedRecover~\cite{cao2022fedrecover} focuses on post-poisoning recovery after an attack, aiming to restore the global model's performance. Ditto~\cite{li2021ditto} integrates fairness and robustness by regularizing the local training objective, and Cronus~\cite{chang2019cronus} enhances security and privacy through knowledge distillation.

\noindent\textbf{Systemizing FL Defenses:}
Besides these few defenses, the variety of available options (Table~\ref{tab:defenses}) poses a challenge for practitioners, i.e., determining the right defense for a specific use case or integrating multiple defenses for enhanced robustness becomes complex without a clear understanding of where a defense and its dependencies fit in the FL pipeline.

To address these research gaps, we conduct a comprehensive systemization (\S\ref{sec:systemization}) of FL defenses, organizing them along three crucial dimensions: processing of client updates, server's knowledge, and defense phase.
While existing works~\cite{shejwalkar2022back, jere2020taxonomy, barreno2010the, biggio2018wild, huang2011adversarial, rodriguez2023survey} have designed taxonomies primarily focused on adversarial ML, including those that guide the selection of the appropriate attacks and settings for FL, a dedicated systemization for FL defenses has been lacking.

To the best of our knowledge, we are the first to propose such a systemization.
Our framework simplifies the selection and integration of defenses, clarifying when and where each defense is applied (\S\ref{systemization:classification}). For example, Figure~\ref{fig:defenses_systemization} shows that FLDetector operates in the pre-aggregation phase(``when") at the server(``where"). Moreover, the systemization highlights underrepresented defense types, encouraging further exploration and innovation. 
For example, 

our analysis identified FedRecover as the only post-training, update-modification technique using estimation, prompting exploration of not only other estimation-based but also post-training defenses. From our systemization in Table~\ref{tab:systemization}, we can see that most defenses are on the server side, operate during pre-aggregation, and employ metric-based processing of client updates. Importantly, our systemization is designed to be expandable to incorporate more attributes in the future.

\noindent\textbf{Pitfalls in experimental setups of FL defenses: }
During our systemization, we find that defenses are evaluated across various experimental setups with different choices of datasets, data distributions, attacks, etc. We thoroughly survey 50 top-tier FL defense works (\S\ref{sec:pitfalls}) and report frequently used choices of six experimental setup components: \emph{data and distribution, FL type and algorithm, attack, and evaluation type}. We find questionable trends in these choices, e.g., most works use the intrinsically robust MNIST dataset and resort to simple attacks such as label flipping. This motivates the need to uncover \emph{pitfalls} in the experimental setups of existing FL defense works and provide guidelines for future evaluations.

Existing literature provides guidelines for robustness evaluations. For instance, \cite{arp2022and} offers insights and recommendations, focusing on pitfalls in centralized ML-based security system evaluations. Similarly, \cite{carlini2019evaluating} identifies pitfalls in evaluating adversarial robustness and suggests mitigation guidelines. In~\cite{shejwalkar2022back}, authors question trends in existing attack threat models, demonstrating FL robustness under practical limitations. However, a comprehensive exploration of FL defenses and their experimental setups is lacking.

To fill this gap in the literature, we identify six distinct pitfalls in the evaluation setups of existing works based on our survey in \S\ref{sec:pitfalls}. We take inspiration from \cite{khan2023pitfalls} in identifying pitfalls in FL defenses and building on top of their work. Not only do we explore additional pitfalls and analyze their impact in much more diverse settings, but we also first perform an extensive systemization to justify the choices we make in pitfall analysis. For example, Figure~\ref{fig:defenses_systemization} and Table~\ref{tab:systemization} help us to select and justify three distinct defenses for the impact analysis of pitfalls in \S\ref{sec:impact}.

Figure~\ref{fig:fl_flowchart} shows the pitfalls at the point of their occurrence in the FL training pipeline. For instance, choosing an intrinsically robust dataset is a pitfall that occurs in the dataset selection stage. We also show the common choices for each stage of the pipeline, such as global or personalized evaluation in the last stage. Figure~\ref{fig:fl_flowchart} also highlights the interdependencies between stages. For example, non-i.i.d. distribution in the distribution stage influences the choice of evaluation metrics (\S\ref{impact:evaluation}), while large-scale limits the threat model in the attack stage (\S\ref{impact:scalability}). We thoroughly explain each pitfall and its prevalence in \S\ref{sec:pitfalls}, and provide actionable recommendations to overcome them. Finally, in \S\ref{sec:impact}, we perform a thorough impact analysis of the identified pitfalls using three representative defenses and show how we can avoid them by following our recommendations.
These pitfalls \emph{can also be applied to attacks}, e.g., only evaluating an attack in cross-silo settings and not considering its efficacy in the cross-device setting is a pitfall. Similarly, other pitfalls can arise from attacks, as attacks and defenses are inherently interconnected—essentially two sides of the same coin working together. \textbf{\emph{However, for the purpose of this paper, we will analyze the pitfalls from the lens of defenses.}}

\noindent In summary, we make the following contributions:

\noindent\textbf{1. Systemization of FL Defenses:} We perform a comprehensive systemization of FL defenses along three dimensions: processing of client updates, server's knowledge, and the defense phase (\S\ref{sec:systemization}).

\noindent\textbf{2. Identifying Major Pitfalls:} We comprehensively review 50 top-tier FL defense works and identify six prevalent pitfalls. In response to each pitfall, we provide actionable recommendations to guide future research efforts (\S\ref{sec:pitfalls}).

\noindent\textbf{3. Dissecting the Impact of Pitfalls:} Guided by our systemization, we choose three representative FL defenses and use them to perform a thorough impact analysis of our identified pitfalls. We show how the pitfalls lead to a false sense of security, and by following our recommendations, the research community can overcome them (\S\ref{sec:impact}).

\begin{mybox}
    This work is not meant as finger-pointing, particularly to the defenses under evaluation. We have chosen them as representative contributions to the field, humbly employing them as testbeds to offer constructive guidelines for future research.
\end{mybox}
\section{Background}\label{sec:background}
    
\subsection{Federated Learning (FL)}\label{background:FL}
In FL~\cite{kairouz2019advances, mcmahan2017communication, konevcny2016federated}, a service provider, called \emph{server}, trains a \emph{global model}, $\theta^g$, on the private data from multiple collaborating clients, all without directly collecting their individual data.
During the $t^{th}$ FL round, the server selects $n$ out of total $N$ clients and shares the most recent global model ($\theta^t_g$) with them. 
Then, a client $k$ uses their local data $D_k$ to compute an update $\nabla^t_k$ and shares it with the server. These updates serve as a client's contribution towards refining a global model. 
Depending on how a client computes their update, FL algorithms can be broadly divided~\cite{mcmahan2017communication} into \emph{FedSGD} and \emph{FedAvg}. 
In \emph{FedSGD}, a client computes the update by sampling a subset $b$ from their local data and calculating a gradient of loss $\ell(b;\theta^t_g)$ of the global model on the subset, i.e., $\nabla^t_k = {\partial \ell(b; \theta^t_g)}/{\partial \theta^t_g}$. In \emph{FedAvg}, a client $k$ \emph{fine-tunes} $\theta^t_g$ on their local data using stochastic gradient descent (SGD) for a fixed number of local epochs $E$, resulting in an updated local model $\theta^t_k$. The client then computes their update as the difference $\nabla^t_k= \theta^t_k-\theta^t_g$ and shares $\nabla^t_k$ with the server.
Next, the server computes an aggregate of client updates using an AGR, $f_\mathsf{agg}$(such as mean), i.e., using $\nabla^t_\mathsf{agg}= f_\mathsf{agr}(\nabla^t_{\{k\in[n]\}})$. Finally, the server updates the global model of the $(t+1)^{th}$ round using SGD as $\theta^{t+1}_g\leftarrow \theta^{t}_g+\eta\nabla^t_\mathsf{agg}$ with server's learning rate $\eta$. Due to these differences, FedAvg achieves faster convergence and attains higher accuracy than FedSGD~\cite{mcmahan2017communication}.

After discussing the update process of FL models and the collaboration between servers and clients, we present the FL applications (\emph{where it is deployed}) and setups (\emph{how it is used}). There are two main types of deployments: \textbf{cross-device} and \textbf{cross-silo}, as explained in~\cite{kairouz2019advances}.
In \emph{cross-device FL}, $N$ is very large (ranging from a few thousand to billions)~\cite{reddi2020adaptive}, and only a tiny fraction of them are chosen in each FL training round ($n\ll N$). These clients are typically resource-constrained devices such as mobile phones, smartwatches, and other IoT devices~\cite{gboard, zhao2022multimodal}. Contrastingly, in \emph{cross-silo FL}, $N$ is moderate (up to 100)~\cite{li2021model}, and all clients are selected in each round ($n=N$). These clients are typically large corporations, including banks~\cite{webankcredit} and hospitals~\cite{nguyen2022federated}.
\subsection{Poisoning Attacks in FL}\label{background:attacks}
\begin{figure}[t]
\centering
\includegraphics[width=0.7\linewidth]{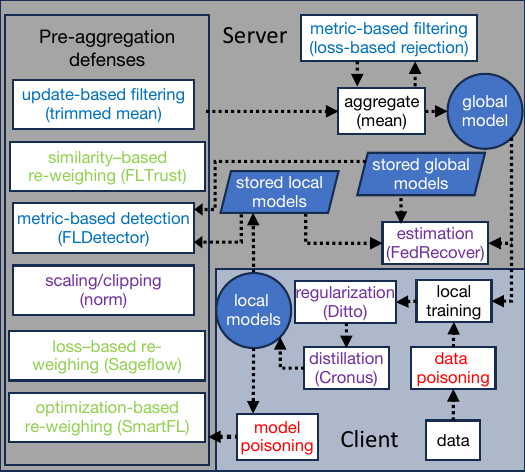}
\caption{Systemizing FL Defenses: Categorization of defenses based on their \emph{defense phases}. Processing operations for client updates are color-coded: filtering (blue), re-weighing (green), and modification (purple). Defenses belonging to the same category of processing may have different phases, e.g., FedRecover performs update modification at the server, while Ditto performs update modification at the client. Dependencies are also highlighted, such as FedRecover utilizing stored local and global models for estimation.}
\label{fig:defenses_systemization}
\end{figure}

\begin{table*}[]
\caption{A systematic overview of FL defenses, helping practitioners in 1) selecting defenses aligned with their use cases, 2) combining multiple defenses for heightened performance, and 3) designing new defense approaches by gaining insights into the FL defenses.}
\label{tab:systemization}
\scriptsize
\resizebox{1\textwidth}{!}{%
\begin{tabular}{|l|l|l|p{0.38\linewidth}|l|}
\hline
\textbf{Dimension} & \textbf{Types} & \textbf{Attributes} & \textbf{Description} & \textbf{Defenses} \\ \hline
\multirow{8}{*}{\textbf{\begin{tabular}[c]{@{}l@{}}Processing \\ of Client \\ Updates\end{tabular}}} & \multirow{2}{*}{Filtering} & Update-based & Filters updates by comparing update values with each other. & TrMean~\cite{yin2018byzantine}, Krum~\cite{blanchard2017machine}, GeometricMean~\cite{pillutla2019robust} \\ \cline{3-5} 
 &  & Metric-based & Filters updates by comparing metrics associated with each update. & MST-AD \& Density-AD~\cite{ranjan2022securing}, LF-Fighter~\cite{jebreel2022defending}, ERR \& LFR~\cite{fang2020local}, FLDetector~\cite{zhang2022fldetector} \\ \cline{2-5} 
 & \multirow{3}{*}{\begin{tabular}[c]{@{}l@{}}Update\\  re-weighing\end{tabular}} & Similarity-based & Filters updates by comparing their similarity with a reference. & CONTRA~\cite{awan2021contra}, FLTrust~\cite{cao2020fltrust} \\ \cline{3-5} 
 &  & Loss-based & Applies weights based on loss for an update. & Sageflow~\cite{park2021sageflow}, Anomaly Detection~\cite{li2020learning} \\ \cline{3-5}
 &  & Optimization-based & Applies weights based on an optimization problem. & SmartFL~\cite{xie2022robust} \\ \cline{2-5} 
 & \multirow{3}{*}{\begin{tabular}[c]{@{}l@{}}Update \\ modification\end{tabular}} & Scaling/Clipping & Scales or clips an update if it exceeds a certain threshold. & Norm Bounding~\cite{sun2019can}, Signguard~\cite{xu2021signguard} \\ \cline{3-5} 
 &  & Distillation & Distills an update into a low-dimensional vector. & Cronus~\cite{chang2019cronus}, Auror~\cite{shen2016auror} \\ \cline{3-5} 
 &  & Regularization & Introduces regularization to control robustness and privacy. & Ditto~\cite{li2021ditto} \\ \cline{3-5} 
 &  & Estimation & Estimates a benign update from historical information. & FedRecover~\cite{cao2022fedrecover} \\ \hline
 \hline
\multirow{4}{*}{\textbf{\begin{tabular}[c]{@{}l@{}}Server's \\ Knowledge\end{tabular}}} & \multirow{2}{*}{\begin{tabular}[c]{@{}l@{}}Knowledge \\ of data\end{tabular}} & No-knowledge & Server has no knowledge of data or its distribution at the client side. & TrMean~\cite{yin2018byzantine}, FedRecover~\cite{cao2022fedrecover}, FLDetector~\cite{zhang2022fldetector}, Cronus~\cite{chang2019cronus} \\ \cline{3-5} 
 &  & Partial knowledge & The server uses a small auxiliary dataset. & Sageflow~\cite{park2021sageflow}, SmartFL~\cite{xie2022robust}, FLTrust~\cite{cao2020fltrust} \\ \cline{2-5} 
 & \multirow{2}{*}{\begin{tabular}[c]{@{}l@{}}Knowledge \\ of updates\end{tabular}} & No-knowledge & Server does not have knowledge of local model updates. & Auror~\cite{shen2016auror}, Cronus~\cite{chang2019cronus} \\ \cline{3-5} 
 &  & Full knowledge & Server has complete knowledge of local model updates. & FLCert~\cite{cao2022flcert}, Krum~\cite{blanchard2017machine}, Anomaly Detection~\cite{li2020learning} \\ \hline
 \hline
\multirow{4}{*}{\textbf{\begin{tabular}[c]{@{}l@{}}Defense \\ Phase\end{tabular}}} & \multirow{2}{*}{Aggregation} & Pre-aggregation & Processing of client updates is done before updates are aggregated. & MST-AD \& Density-AD~\cite{ranjan2022securing}, Bulyan~\cite{mhamdi2018the}, TrMean~\cite{yin2018byzantine}, Krum~\cite{blanchard2017machine}, Sageflow~\cite{park2021sageflow} \\ \cline{3-5} 
 &  & Post-aggregation & Processing of client updates is done after aggregation. & FLCert~\cite{cao2022flcert}, ERR \& LFR~\cite{fang2020local} \\ \cline{2-5} 
 & \multirow{2}{*}{\begin{tabular}[c]{@{}l@{}}Non-\\ aggregation\end{tabular}} & Local training & The defense component is part of the local training. & FLIP~\cite{zhang2022flip}, Ditto~\cite{li2021ditto}, Cronus~\cite{chang2019cronus} \\ \cline{3-5} 
 &  & Post-training & The defense is applied after training. & FedRecover~\cite{cao2022fedrecover} \\ \hline
\end{tabular}}
\end{table*}
There are various poisoning attacks in literature~\cite{blanchard2017machine, baruch2019a, bhagoji2019analyzing, bagdasaryan2018how, mhamdi2018the, fang2020local, mahloujifar2019universal, xie2019fall, munoz2017towards, shejwalkar2021manipulating}.
An \emph{untargeted} poisoning attack aims to lower the test accuracy for all test inputs indiscriminately~\cite{fang2020local, baruch2019a, mhamdi2018the, mahloujifar2019universal, xie2019fall}. A \emph{targeted} poisoning attack~\cite{bhagoji2019analyzing, bagdasaryan2018how} lowers the accuracy on specific test inputs.
For instance, in \emph{backdoor attacks}~\cite{bagdasaryan2018how} (a sub-category of targeted attacks), the goal is to misclassify only those test inputs that have an embedded \emph{backdoor trigger}. Since these attacks only affect a subset of inputs, they are much weaker than untargeted attacks.

We can also divide the attacks based on the adversary's capabilities. In model poisoning attacks~\cite{fang2020local, baruch2019a, mhamdi2018the, xie2019fall, bhagoji2019analyzing, bagdasaryan2018how, shejwalkar2021manipulating}, the adversary is strong enough to access and perturb the model gradients on malicious devices before they are sent to the server in every training round. A data-poisoning adversary~\cite{munoz2017towards} is much weaker than the model poisoning adversary~\cite{baruch2019a, fang2020local, shejwalkar2022back, shejwalkar2021manipulating} as it can only poison the datasets on malicious devices. 
\subsubsection{Attacks used in our Study}\label{background:attacks_study}
In our evaluation, we focus on untargeted model poisoning attacks \emph{as they are stronger}~\cite{shejwalkar2022back}.

\noindent\textbf{Stat-Opt~\cite{fang2020local}:} provides a generic model poisoning method and tailors it to specific AGRs such as TrMean~\cite{yin2018byzantine}, Median~\cite{yin2018byzantine}, and Krum~\cite{blanchard2017machine}. The adversary first calculates the mean of the benign clients' updates, $\nabla^b$, and finds the \emph{static} malicious direction $w = -sign(\nabla^b)$. 
It directs the benign average along the calculated direction and scales it with $\gamma$ to obtain the final poisoned update, $-\gamma w$.

\noindent\textbf{Dyn-Opt~\cite{shejwalkar2021manipulating}:} proposes a general poisoning method and tailors it to specific AGRs, similar to Stat-Opt. The key distinction lies in the \emph{dynamic} and \emph{data-dependent} nature of the perturbation. The attack starts by computing the mean of benign updates, 
$\nabla^b$, and a data-dependent direction, $w$. The final poisoned update is calculated as $\nabla^` = \nabla^b + \gamma w$, where the attack finds the largest $\gamma$ that can bypass the AGR. They compare their attack with Stat-Opt and show that the dataset-tailored $w$ and optimization-based scaling factor $\gamma$ make their attack a much stronger one.
\subsubsection{Threat Model in Our Study}\label{background:threat_model}
Below, we detail the threat models for poisoning attacks used in our study.

\noindent\textbf{Goal:}
Our adversary is \emph{untargeted}, crafts malicious updates and sends them to the server, where, upon aggregation with other updates, it indiscriminately lowers the accuracy of the global model for all test inputs.

\noindent\textbf{Knowledge:} In line with most defense approaches, we assume that the adversary knows the AGR used by the server.
Unless explicitly specified, the adversary has complete knowledge of the gradients of both malicious and benign clients. In some cases, we employ a partial knowledge adversary, where the adversary knows the server's AGR but not the gradients of benign clients.

\noindent\textbf{Capabilities:} Our untargeted model poisoning adversary controls $m$ out of $N$ clients. The adversary is strong enough to manipulate model updates of the malicious clients it controls and has access to the global model parameters shared every round. 
We set the proportion of malicious clients at 20\% (unless stated otherwise), a common benchmark in prior studies~\cite{shejwalkar2021manipulating, fang2020local, cao2020fltrust, cao2022fedrecover}, which also examined how varying this percentage impacts the severity of attacks. To ensure consistency and comparability with these works, we adhere to the same 20\% setting in our implementation.
\section{Systemization of Defenses Against  FL Poisoning}\label{sec:systemization}
Here, we introduce a systematization for FL defenses and use it to
rationalize the selection of three representative defenses from the literature. Later, 
in \S\ref{sec:impact}, we use these chosen defenses to conduct a comprehensive impact analysis of the pitfalls outlined in \S\ref{sec:pitfalls}.
\subsection{Classification of Defenses}\label{systemization:classification}
Here, we present the three key dimensions along which we classify FL defenses in literature, as shown in Table~\ref{tab:systemization}. In Figure~\ref{fig:defenses_systemization}, we group defenses according to the third dimension, i.e., \emph{defense phase}, and highlight their dependencies. 

\subsubsection{Processing of client updates}\label{systemization:classification:operation}
Client model updates undergo several processing steps before they are aggregated at the server. The commonly used processing operations are:

\noindent\textbf{Filtering updates} 
to entirely or partially eliminate local updates. \emph{Filtering} defenses fall into two main categories: 
those based on the values of local model updates, termed \emph{update-based filtering}~\cite{yin2018byzantine, blanchard2017machine, pillutla2019robust}, and those relying on some metrics associated with local models, known as \emph{metric-based filtering}~\cite{ranjan2022securing, jebreel2022defending, fang2020local, zhang2022fldetector}.

\begin{enumerate}[leftmargin=*, label=\alph*), wide]
    \item \emph{Update-based filtering ~\cite{yin2018byzantine, blanchard2017machine, pillutla2019robust}} is based on the values of local model updates. It further divides into \emph{dimension-wise} and \emph{vector-wise} filtering. Dimension-wise filtering defenses, such as TrMean~\cite{yin2018byzantine} and Median~\cite{yin2018byzantine}, filter out malicious values along each update dimension, while vector-wise filtering defenses, such as, Krum~\cite{blanchard2017machine}, remove entire malicious updates.
    \item \emph{Metric-based filtering~\cite{ranjan2022securing, jebreel2022defending, fang2020local, zhang2022fldetector}} relies on some metrics associated with local models, e.g., FLDetector~\cite{zhang2022fldetector} uses a \emph{suspicious score} as the metric to identify and remove malicious clients from the training process. We describe FLDetector in detail in \S\ref{systemization:defenses_study:descriptions}. In \emph{loss-based rejection}~\cite{fang2020local}, the loss associated with and without incorporating an update for aggregation is calculated, and updates with higher loss are removed. Similarly, \emph{error-based rejection}~\cite{fang2020local} removes updates by assessing the error instead of the loss.
\end{enumerate}

\noindent\textbf{Update re-weighing} 
involves assigning a weight to each local update, reflecting its perceived level of maliciousness. Various re-weighting approaches exist, including:
\begin{enumerate}[leftmargin=*, label=\alph*), wide]
    \item \emph{Similarity-based re-weighting~\cite{awan2021contra, cao2020fltrust}} is shown by FLTrust~\cite{cao2020fltrust}, where the server assigns a trust score to each client based on the similarity of its updates to the server's update, computed on a small dataset.
    \item \emph{Loss-based re-weighting~\cite{park2021sageflow, li2020learning}} illustrated by Sageflow~\cite{park2021sageflow}, involves the server assigning a weight to each update based on local model loss on a small dataset at the server.
    \item \emph{Optimization-based re-weighting} is demonstrated by SmartFL~\cite{xie2022robust}, which assigns weights through an optimization problem with the same number of parameters as that of clients.
\end{enumerate}
\noindent\textbf{Update Modification} changes the update itself to safeguard the global model from the impact of malicious updates. The key techniques within this approach are:
\begin{enumerate}[leftmargin=*, label=\alph*), wide]
    \item \emph{Scaling Updates} limits a local update by clipping it if it exceeds a certain threshold, for example, defenses like Norm-bounding~\cite{sun2019can, xu2021signguard}.
    \item \emph{Distilling update knowledge~\cite{chang2019cronus, shen2016auror}} is another facet of update modification that involves avoiding the transmission of the entire local model update to the server due to the \emph{curse of dimensionality}~\cite{chang2019cronus}, which increases the risk of higher impact from poisoning attacks. Instead, clients send distilled information to the server. In Cronus~\cite{chang2019cronus}, clients send soft labels to the server, and the aggregate is used to update local models. This defense mitigates the risk of poisoning attacks and directly prevents whitebox inference attacks, as the server cannot access the local model parameters.
    \item \emph{Regularization} defenses perform regularization to achieve personalized~\cite{hanzely2020federated, hanzely2020lower} client models. An example is Ditto~\cite{li2021ditto}, which modifies the local training objective by introducing a regularization term to control the tradeoff between privacy and robustness of the local model.
    \item \emph{Estimation} is exemplified by FedRecover~\cite{cao2022fedrecover}, which leverages past information and estimation to recover the unpoisoned global model. We discuss the mechanism of FedRecover in detail in \S\ref{systemization:defenses_study:descriptions}.
\end{enumerate}
\subsubsection{Server's Knowledge}\label{systemization:classification:knowledge}
The defense is generally applied at the server, where it collects all local updates and strives to obtain the best possible global model that is least affected by malicious updates. The server's knowledge varies across different defenses and can be described in terms of data and local model updates:

\noindent\textbf{Knowledge of Data:} 
In the \emph{no-knowledge} setting~\cite{yin2018byzantine, cao2022fedrecover, zhang2022fldetector, chang2019cronus}, as the name suggests, the server lacks information about the data used for training and testing. It only possesses knowledge of the collected local model updates, examples of which include TrMean~\cite{yin2018byzantine}, Krum~\cite{blanchard2017machine}, and Median~\cite{yin2018byzantine}. Conversely, in the \emph{partial-knowledge} setting~\cite{park2021sageflow, xie2022robust, cao2020fltrust}, the server possesses a small dataset, which it deploys in various ways, such as calculating entropy associated with updates~\cite{park2021sageflow} or assigning a trust score to each client~\cite{cao2020fltrust} to enhance aggregation robustness against attacks.
\begin{figure*}[t]
\centering
\includegraphics[width=.97\linewidth]{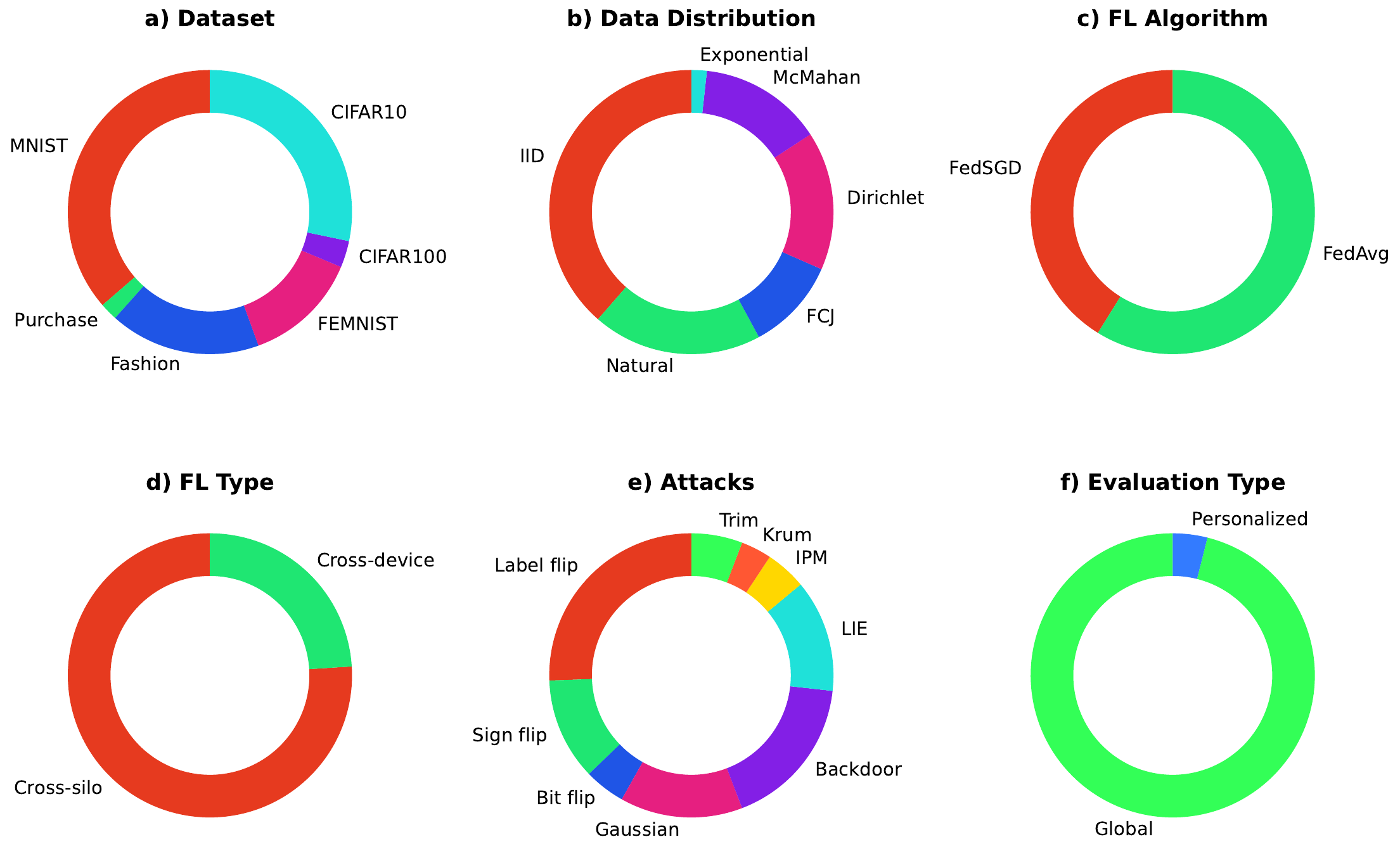}
\caption{Frequency of choices of the six key components of robustness evaluation setup: dataset, distribution of clients' data, FL algorithm, FL type, attacks, and evaluation. \S~\ref{sec:impact} discusses the impacts of choices on the robustness of FL poisoning defenses.}
\label{fig:defense_survey}
\end{figure*}

\noindent\textbf{Knowledge of Local Model Updates:} 
In the \emph{full-knowledge} setting~\cite{cao2022flcert, blanchard2017machine, li2020learning}, the server has complete access to the model parameters of all clients, representing the widely used scenario. In the \emph{partial-knowledge}, or \emph{distilled-knowledge} setting~\cite{shen2016auror, chang2019cronus}, the server only has access to some distilled form of the model parameters, such as the output layers~\cite{chang2019cronus}.
\subsubsection{Defense Phase}\label{systemization:classification:phase}
We categorize defenses based on the training \emph{phase}, specifying when and where in the training pipeline the defense is applied.

\noindent\textbf{Aggregation-based} defenses are applied during the aggregation phase. These defenses can be further categorized into \emph{pre-aggregation} \cite{ranjan2022securing, mhamdi2018the, yin2018byzantine, blanchard2017machine, park2021sageflow} defenses that perform processing of client updates such as dimension-wise filtering before aggregating updates, or \emph{post-aggregation}~\cite{cao2022flcert, fang2020local, cao2021provably} defenses, e.g., Ensemble FL~\cite{cao2021provably} that creates all possible aggregations of k models from N clients, then selects the most frequent predicted label as the correct one.

\noindent\textbf{Non-aggregation-based} defenses that are not performed at aggregation can be further categorized based on their \emph{phase}:
\begin{enumerate}[leftmargin=*, label=\alph*), wide]
    \item \emph{During Local Training:}~\cite{zhang2022flip, li2021ditto, chang2019cronus} The defense takes place at the client's side during local training. For instance, Ditto~\cite{li2021ditto} uses regularization in client training to control the deviation of benign local models from poisoned global models.
    \item \emph{Post-Training:} FedRecover~\cite{cao2022fedrecover}, employing a recovery-based mechanism, falls into this category as it requires historical information from one training session to estimate the un-poisoned global model during the \emph{recovery} phase.
\end{enumerate}

\subsection{Our Selected Case Study Defenses}\label{systemization:justification}

We choose three defenses for brevity to analyze the impact of pitfalls in \S\ref{sec:impact}; TrMean, FLDetector, and FedRecover. We first provide each of their descriptions and then give a detailed justification for choosing them for our impact analysis.

\subsubsection{Descriptions of our chosen defenses}\label{systemization:defenses_study:descriptions}
\paragraphb{Trimmed Mean (\emph{TrMean})}~\cite{yin2018byzantine}
\label{background:trmean}
is a foundational defense used in advanced defenses~\cite{cao2022fedrecover,zhang2022fldetector,shejwalkar2021manipulating}. It sorts each input dimension $j$ of the client updates, discards the $m$ largest and smallest values (where $m$ indicates compromised clients), and averages the rest.

\paragraphb{FLDetector~\cite{zhang2022fldetector}}\label{background:fld}
is designed to \emph{detect} and eliminate malicious clients, ensuring a byzantine-robust FL system obtains a precise global model. FLDetector operates on the principle that malicious updates, tainted by adversaries, differ statistically from benign ones.

For discerning these updates, the server estimates a global model update for client $k$ at round $t$ using the L-BFGS algorithm: ($\hat{\nabla}_{t}^{k} = \nabla_{t-1}^{k} + \hat{H}^{t}\cdot(\theta_{t} - \theta_{t-1})$). Here, $\nabla_{t-1}^{k}$. The server retains past $N$ global model differences ($\Delta \theta_{t}$) and updates differences ($\Delta \nabla_{t-1}$) to compute the \emph{HVP (Hessian Vector Product)} with \emph{L-BFGS}. It then gauges a client's \emph{suspicious score} by comparing actual and predicted updates through their Euclidean distance. With scores from the last $N$ rounds, clients are clustered via Gap statistics~\cite{tibshirani2001estimating} and K-means. The group with higher average scores is deemed malicious. Upon detecting a rogue client, the server ceases training, removes the offender, and restarts training to achieve enhanced accuracy.

\paragraphb{FedRecover~\cite{cao2022fedrecover}}\label{background:fdr}
aims to \emph{recover} an FL global model compromised by a poisoning attack. In the \emph{original training phase}, for each round $t$, FedRecover saves $\nabla^k_t$ from client $k$ and global models $\theta^t_g$. This data is used as \emph{historical information} in the \emph{recovery phase}, which consists of the following stages.
In the \emph{warmup phase}, the server requests clients' \emph{exact updates} for the initial $T_w$ rounds. In the \emph{estimation phase}, the server \emph{estimates} client updates each round, with $\hat{\nabla}^t_k$\ representing client $k$'s estimated update at round $t$. This estimate is derived using the L-BFGS algorithm~\cite{nocedal1980updating} based on the original global model, client update, and recovered global model. The model's estimated update is defined as $\hat{\nabla}^k_t = \overline{\nabla}^t_k + \Tilde{H}^t_k(\hat{\theta}^t_g - \overline{\theta}^t_g)$, where the latter term is the HVP(Hessian Vector Product).
Every $T_c$ rounds, to ensure the estimated global model $\hat{\theta}^t_g$ aligns closely with the accurate model, the server initiates a \emph{periodic correction} by requesting \emph{exact updates}. If any component of a client's estimated update surpasses the \emph{abnormality threshold} $\tau$, that client is prompted for an exact update.
In the final \emph{fine-tuning phase}, spanning $T_f$ rounds, clients are asked to provide their exact updates,$\nabla^t_k$, to refine the global model by eliminating potential estimation errors.

\subsubsection{Justification for our choice of representative defenses}\label{systemization:defenses_study:justification}
We justify that the chosen defenses are distinct along the dimensions in Table~\ref{tab:systemization}.
TrMean~\cite{yin2018byzantine} is a \textbf{\emph{pre-aggregation, update-based filtering}} defense that removes malicious components of client updates dimension-wise during training. FLDetector~\cite{zhang2022fldetector} is a \textbf{\emph{pre-aggregation, metric-based filtering}} defense designed to detect and remove malicious clients during training by analyzing their updates. The metric FLDetector uses is the \emph{suspicious score}, which measures the consistency between the actual update and an estimated one. It is important to note that although TrMean and FLDetector seem similar in their filtering mechanisms, they are different and have distinct approaches. TrMean filters \emph{components} of updates before aggregating them, while FLDetector removes entire \emph{clients} from the training process if they deviate too much from an estimated reference update, which is calculated using historical information. FedRecover~\cite{cao2022fedrecover} is a \textbf{\emph{post-training, update modification}} defense that uses \emph{estimation} to \emph{recover} from a previously poisoned global model. It is an advanced defense that uses TrMean in its aggregation phase and relies on a detection mechanism to remove malicious clients before it starts the recovery process. In \S\ref{impact:datasets:fdr} and \S\ref{impact:attacks:fdr}, we show the performance of FedRecover when we perform Stat-Opt on TrMean and FLDetector, respectively.

All three defenses do not require any auxiliary dataset at the server, and we prefer this setting because the server might not have access to the dataset in practical, real-world scenarios. Similarly, we prefer the \emph{full-knowledge} of client updates scenario, as most of the works~\cite{cao2022flcert, blanchard2017machine, li2020learning, li2021ditto, park2021sageflow, cao2020fltrust} use this setting, and it leads to a stronger adversarial setting. Although the server in all three of our chosen defenses has full knowledge of the client model updates, the defenses differ in the amount of historical information needed. Figure~\ref{fig:defenses_systemization} clearly shows these different dependencies. TrMean does not require model updates from the past rounds and performs filtering using client model updates of the current round. FLDetector requires updates from the past \emph{few} rounds and uses them to calculate the malicious score for updates in the current round. Since FedRecover is a post-training defense, it requires updates from all the rounds in the original training. Therefore, the volume of updates required is highest for FedRecover and lowest for TrMean.

\section{Pitfalls in FL Defense Evaluations}\label{sec:pitfalls}
After presenting the detailed systemization of defenses, it is imperative to unveil critical pitfalls in FL robustness evaluations. By scrutinizing 50 defenses (Table~\ref{tab:defenses} in Appendix), we link each pitfall to specific components in the FL workflow (Figure~\ref{fig:fl_flowchart}). We examine each pitfall's prevalence (Figure~\ref{fig:defense_survey}) across the 50 works, discuss their implications, and conclude with practical recommendations. 

\begin{mybox}
\paragraphb{Pitfall-1: Intrinsically robust datasets.}\label{pitfalls:1}
The chosen datasets are intrinsically robust and lead to incorrect conclusions about a defense's performance.
\end{mybox}
\noindent\textbf{Description:}
A designed defense may seem to perform well against specific attacks upon evaluation~\cite{yin2018byzantine}. However, the evaluation dataset might be inherently robust because it is simple and lacks complexity. Therefore, in such situations, we cannot tell if an attack is mitigated due to the inherent robustness of the dataset or the effectiveness of the defense.

\noindent\textbf{Prevalence and implications:}
While it is intuitive that a defense mechanism's performance inevitably varies across datasets, using overly simple datasets like MNIST fails to yield meaningful insights into the true robustness of a defense mechanism, (\S\ref{impact:datasets}). MNIST is a class-balanced dataset used in FL using synthetic techniques such as Dirichlet Distribution~\cite{minka2000estimating}. Conversely, real-world FL tasks are complex and characterized by highly class-imbalanced datasets (\S\ref{impact:distribution:statistical_analyses}).

Despite efforts to create open-source datasets mirroring real-world scenarios~\cite{caldas2018leaf, du2022flamby}, \textbf{\emph{our survey reveals that MNIST remains predominant, constituting 30\% of the  works}~\cite{karimireddy2020byzantine,yin2018byzantine,cao2021provably,wu2020mitigating,li2019rsa,liu2021towards}} (Figure~\ref{fig:defense_survey}a). CIFAR10 and FashionMNIST, though commonly used, lack true FL representation due to their class-balanced nature (\S\ref{impact:distribution:statistical_analyses}). FEMNIST~\cite{caldas2018leaf}, a real-world dataset specifically curated for FL, is used by only $20\%$.
Another interesting observation from Figure~\ref{fig:defense_survey}a is the \emph{exclusive reliance on image-classification datasets}, despite the popularity of language and vision-language models in contemporary research.

\noindent\textbf{Recommendations:}
Future Evaluations should consider using FL tasks of varying complexities, such as FEMNIST and CIFAR10, for classification and exploring other modalities, such as language, for NLP tasks. In our evaluations in \S\ref{sec:impact} we use image classification datasets, FEMNIST and CIFAR10, 
and a language dataset, StackOverflow, in \S\ref{impact:scalability:stackoverflow} for large-scale FL evaluation.
\begin{mybox}
\paragraphb{Pitfall-2: Homogeneous client data distributions.}\label{pitfalls:2}
Using i.i.d. (homogeneous) distributions with low heterogeneity may create a deceptive sense of system robustness that does not reflect real-world complexities.
\end{mybox}
\noindent\textbf{Description:}
Evaluating a defense on a particular dataset may yield perceived robustness due to inherent homogeneity (\S\ref{impact:distribution:statistical_analyses}) in the dataset distribution rather than the efficacy of the defense technique itself. 

\noindent\textbf{Prevalence and implications:}
We find that \textbf{\emph{around 50\% of the works, use i.i.d. distributed data, despite evidence that it is easier to defend against such distributions}}~\cite{fang2020local,shejwalkar2021manipulating,baruch2019a,zawad2021curse} (as seen in Figure~\ref{fig:defense_survey}b). 
The second most common approach involves a natural data distribution where each sample is associated with a client such as StackOverflow~\cite{stackoverflow}.
Other artificial distributions include \emph{FCJ}~\cite{fang2020local}, \emph{Dirichlet (Dir)}~\cite{bagdasaryan2018how,reddi2020adaptive} and \emph{Mcmahan}~\cite{mcmahan2017communication}.
In \S\ref{impact:distribution:statistical_analyses}, we prove that Dirichlet more closely aligns with real-world distributions, informing our subsequent analysis in \S\ref{impact:distribution} on how defense performance varies with different distributions and levels of heterogeneity.

\noindent\textbf{Recommendations:}
Future works should prioritize the use of real-world datasets to provide a more realistic evaluation. When working with class-balanced datasets like MNIST, FashionMNIST, and CIFAR10, it is crucial to distribute them among clients heterogeneously. This approach aims to mimic, to some extent, the diversity found in real-world distributions.
\begin{mybox}
\paragraphb{Pitfall-3: Slow-converging algorithms.}\label{pitfalls:3}
Evaluations often overlook the use of state-of-the-art, fast-converging algorithms, thereby compromising robustness.
\end{mybox}
\noindent\textbf{Description:}
The robustness evaluation of an FL defense may heavily rely on the choice of FL algorithms, such as FedAvg or FedSGD, used in its implementation. 
Using a superior algorithm can enhance system robustness by addressing potential weaknesses in the FL algorithm rather than focusing solely on defense improvement. 

\noindent\textbf{Prevalence and implications:} 
Despite FedAvg's recognized advantages in performance, faster convergence, and lower communication overhead compared to FedSGD~\cite{mcmahan2017communication}, \textbf{\emph{approximately 40\% of prior works employ the slow-converging FedSGD algorithm for evaluations}} (Figure~\ref{fig:defense_survey}c). This suboptimal choice contributes to a larger window for attacks and results in a more significant accuracy drop (\S\ref{impact:algorithm}).

\noindent\textbf{Recommendations:}
Future evaluations should prioritize state-of-the-art, fast-converging FL algorithms to remove any weaknesses (such as slow convergence) associated with the FL algorithm.
\begin{mybox}
\paragraphb{Pitfall-4: Limited FL settings.}\label{pitfalls:4}
Considering only limited scenarios while ignoring practical limitations and factors related to scalability, such as computation, communication, cost, and storage.
\end{mybox}
\noindent\textbf{Description:}
The performance of a defense can vary when constrained by real-world limitations, e.g., cost constraints may lead to selecting a very low percentage of malicious clients. Also, the computation, communication, cost, and storage overhead associated with scaling up the system might not be feasible in practical scenarios.

\noindent\textbf{Prevalence and implications:} 
Only $24\%$ of the works in our survey use the cross-device setting (Figure~\ref{fig:defense_survey}d).
As demonstrated by~\cite{shejwalkar2022back} on FEMNIST, CIFAR10, and Purchase datasets with the cross-device setting, using a low percentage of malicious clients due to cost constraints reduces attack performance. We show that on an even larger scale using the naturally distributed Stackoverflow dataset in the cross-device setting, the attack shows \emph{shows no effect on the non-robust mean AGR} (\S\ref{impact:scalability:stackoverflow}). Additionally, defenses that rely on consistent historical information~\cite{zhang2022fldetector, cao2022fedrecover} are incompatible with the cross-device setting (\S\ref{impact:scalability:fdr_fld}) because a client is not selected in every round. We thoroughly discuss the scalability issues of FedRecover and FLDetector in \S\ref{impact:scalability:fdr_fld}.

\noindent\textbf{Recommendations:}
Future evaluations should include deployment conditions of a much larger scale (cross-device) and ensure that the defense provides significant utility compared to the computation and communication cost it incurs.
While we do not label the cross-silo setting as impractical, we emphasize designing defenses that are compatible with the cross-device setting as well. 

\begin{mybox}
\paragraphb{Pitfall-5: Naive attacks.}\label{pitfalls:5}
Evaluating defenses solely against simple and naive attacks rather than incorporating strong state-of-the-art attacks makes a defense seem robust.
\end{mybox}
\noindent\textbf{Description:} 
The true robustness of a defense emerges when tested against strong and adaptive attacks, i.e., attacks tailored for a defense algorithm. Relying on attacks known to be weak and naive for evaluating a new defense will not give us a true picture of the defense's robustness.

\noindent\textbf{Prevalence and implications:} 
Figure~\ref{fig:defense_survey}e shows the frequency of various attacks used in our survey.
Despite the existence of several strong poisoning attacks in the literature~\cite{shejwalkar2021manipulating,fang2020local,xie2019fall,baruch2019a, wang2020attack}, our analysis reveals that \textbf{\emph{about 40\% of the works opt for simplistic approaches}} such as random Gaussian~\cite{blanchard2017machine}, label flipping~\cite{li2021ditto}, sign flipping~\cite{li2019rsa}, bit flipping~\cite{xie2018generalized}), even though prior works have shown their poor performance even under strong adversarial settings~\cite{shejwalkar2022back,fang2020local,li2019rsa,praneeth2020learning}.
We discuss the impact of this pitfall extensively in~\S\ref{impact:attacks} by using state-of-the-art attacks, and in \S\ref{impact:attacks:fld}, we design our own adaptive attack against FLDetector. Our impact analysis reveals that this is one of the most crucial components of FL evaluation.

\noindent\textbf{Recommendations:}
To correctly evaluate the robustness of a defense, one should use 1) strong adaptive attacks that are tailored to the defense algorithm and maximally reduce its performances and 2) strong state-of-the-art attacks from existing literature.
\begin{figure*}[t]
    \begin{subfigure}{0.24\textwidth}
        \centering
    \includegraphics[width=1.05\linewidth]{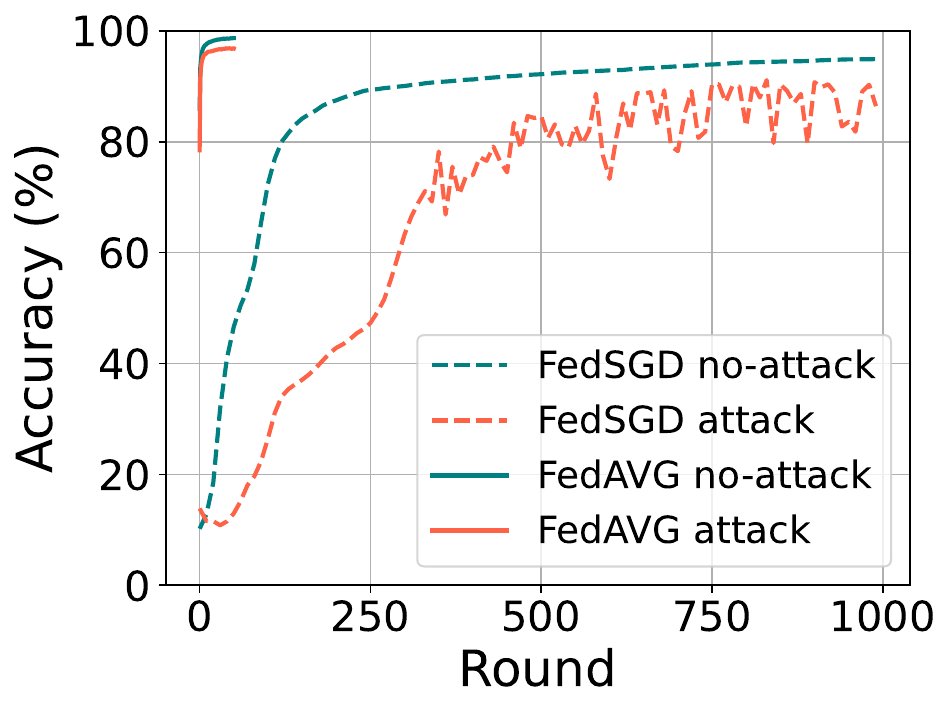}
        \caption{MNIST}
        \label{fig:baseline_mnist}
    \end{subfigure}
    \hfill
    \begin{subfigure}{0.24\textwidth}
        \centering
        \includegraphics[width=\linewidth]{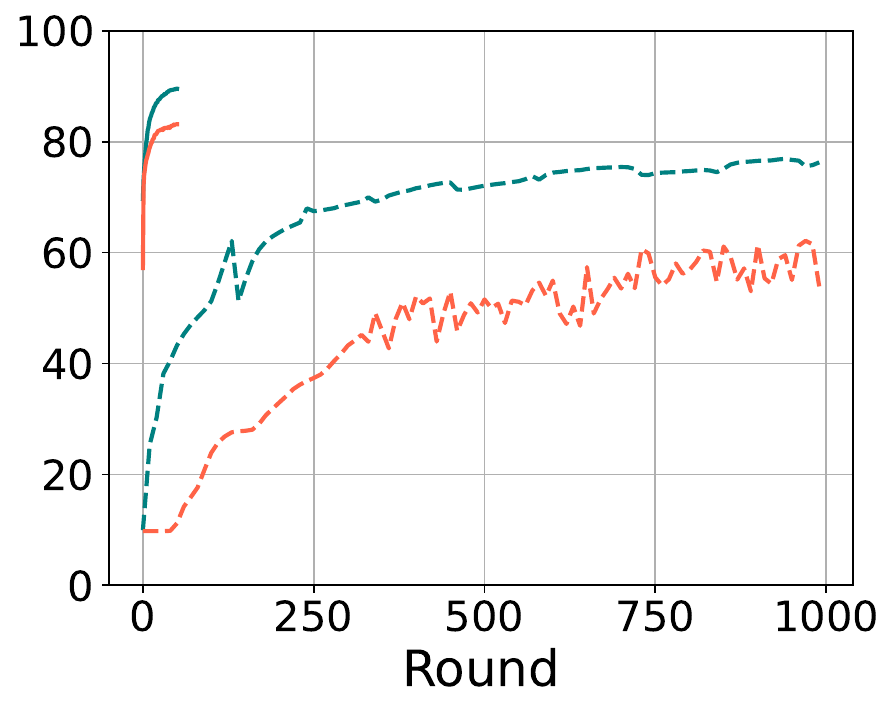}
        \caption{FashionMNIST}
        \label{fig:baseline_fashion}
    \end{subfigure}
    \begin{subfigure}{0.24\textwidth}
        \centering
        \includegraphics[width=\linewidth]{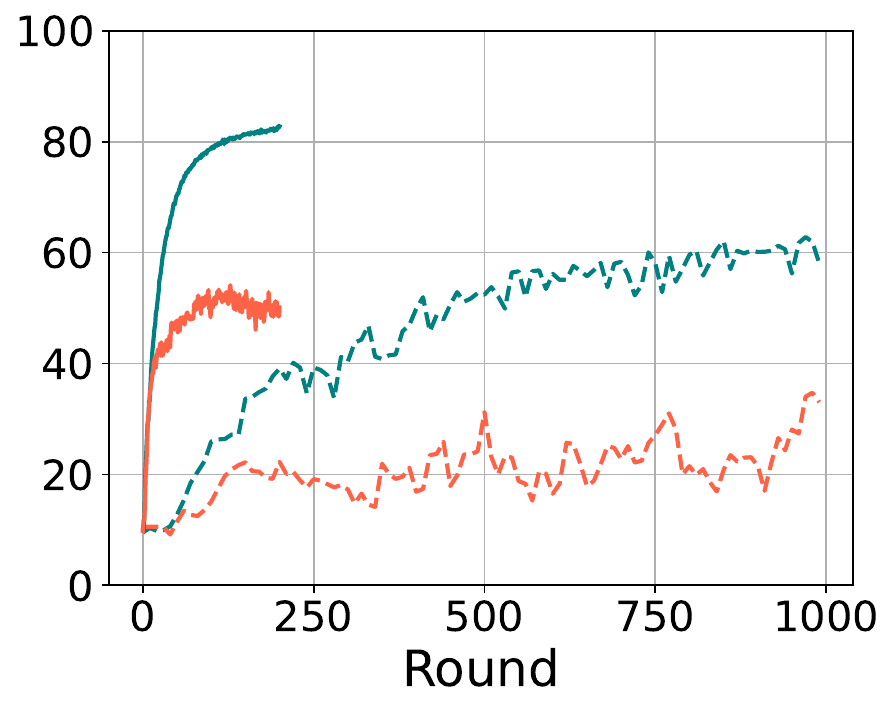}
        \caption{CIFAR10}
        \label{fig:baseline_cifar}
    \end{subfigure}
    \begin{subfigure}{0.24\textwidth}
        \centering
        \includegraphics[width=\linewidth]{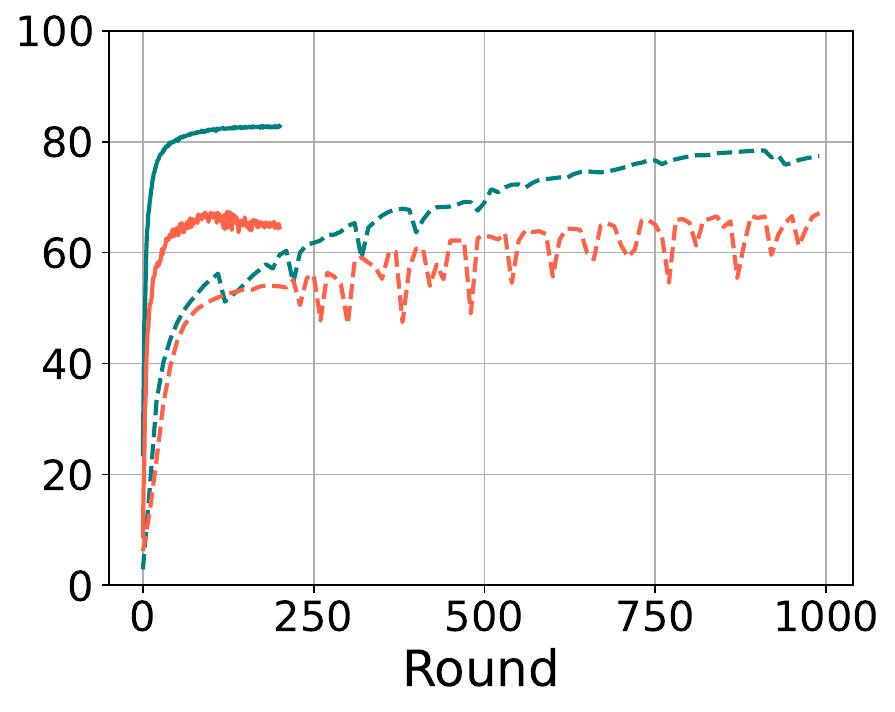}
        \caption{FEMNIST}
        \label{fig:baseline_femnist}
    \end{subfigure}
    \caption{Comparative analysis of TrMean AGR with FedSGD and FedAvg under trim attack. TrMean is more susceptible to poisoning with FedSGD due to slow convergence, which gives adversaries more time for poisoning (threat model in~\S\ref{background:threat_model})}
\label{fig:fl_baseline}
\end{figure*}
\begin{mybox}
\paragraphb{Pitfall-6: Unfair metrics.}\label{pitfalls:6}
Not capturing clients' performances separately and only reporting global accuracy metrics does not give a good measure of per-client robustness.
\end{mybox}
\noindent\textbf{Description:}
Data heterogeneity across clients in practical FL systems results in varying performance across clients. This phenomenon is also known as \textit{representation disparity}~\cite{hashimoto2018fairness}. Global accuracy does not give us any idea of the individual performances of clients.
In addition to this, real-world datasets are class imbalanced~\cite{caldas2018leaf} as opposed to synthetic datasets such as MNIST and CIFAR10. The commonly reported overall accuracy metric lacks information on per-class performance.\\
\noindent\textbf{Prevalence and implications:}
Despite heterogeneity being a well-known issue, only $4\%$ of the surveyed works incorporate personalized evaluations (Figure~\ref{fig:defense_survey}f).
Our per-client analysis of TrMean and FedRecover demonstrates that per-client performances vary a lot, highlighting the need to account for these variations in real-world FL systems (\S\ref{impact:evaluation}). We also highlight the importance of reporting per-class accuracies, as we show that the overall accuracy is a misleading metric (\S\ref{impact:evaluation:imbalance}).

\noindent\textbf{Recommendations:}
Future evaluations should include personalized evaluations along with global evaluations to capture the variation in clients' performances. In addition to this per-client evaluation, future works should also report per-class performance, especially when using class-imbalanced datasets.

\section{Experimental setup}\label{sec:setup}
In this section, we provide the details of our experimental setup.

\subsection{FEMNIST~\cite{caldas2018leaf,cohen2017emnist}}\label{setup:femnist}
\begin{figure}[t]
\centering
\includegraphics[width=0.5\linewidth]{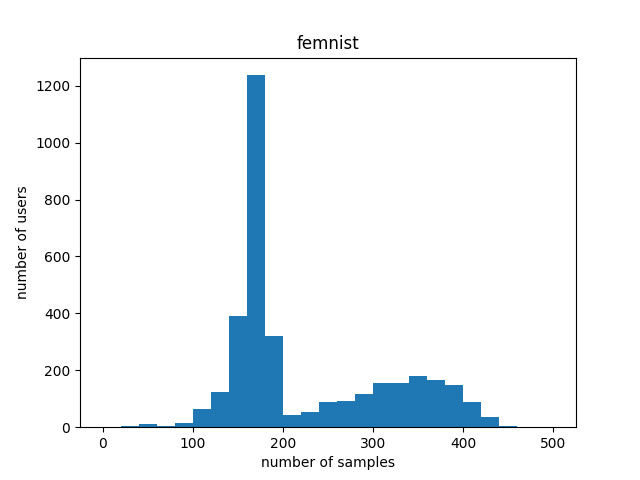}
\vspace{-0.55cm}
\caption{FEMNIST histogram from leaf.cmu.edu}
\label{fig:femnist_hist}
\vspace{-0.3cm}
\end{figure}

FEMNIST (Figure~\ref{fig:femnist_hist}) is a character recognition classification task with 3,400 clients, 62 classes (52 for upper and lower case letters and 10 for digits), and 671,585  grayscale images. 
Each client has data of their own handwritten digits or letters. We use 300 randomly selected clients with their original data in a cross-silo fashion, as FedRecover uses the cross-silo setting in its implementation.
We use the CNN used by ~\cite{cao2022fedrecover} and use the Xavier weight initialization.

\noindent\textbf{Hyperparameters for re-eval:} For FEMNIST, we run over 200 epochs with 300 clients. In the attack setting, 60 clients are malicious. The results in Figure~\ref{fig:fdr_recovery_benign_attack} use $T_w=10$, and $T_c=10$. The FL algorithm used here is FedAVG with a local learning rate of $0.05$ and a global learning rate of $1$. We keep the batch size to 32. The number of local epochs is kept at $1$. For Figure~\ref{fig:fedrec_fnr_fpr_twarm20}, we consider the possibility of benign clients being misclassified as malicious or malicious clients being misclassified as benign, so we vary the false negative and false positive rates between 0.1 and 0.5. 

\subsection{CIFAR10~\cite{Krizhevsky2009learning}}\label{setup:cifar10}
CIFAR10 is a 10-class classification task with 60,000 total RGB images, each of size 32 $\times$ 32. We divide all the data among 100 clients using either Dirichlet~\cite{reddi2020adaptive} or \emph{FCJ}~\cite{fang2020local} distributions, which are the two most popular synthetic strategies to generate the FL dataset. We use a Resnet20 model with the CIFAR dataset.

\noindent\textbf{Hyperparameters for re-eval:} We run over 100 epochs with 100 clients. In the attack setting, 20 clients are malicious. The FL algorithm used here is FedAVG, with a local learning rate of 0.01 and a global learning rate of 1. We keep the batch size to 16. The number of local epochs is kept at $2$. The results in Figure~\ref{fig:fdr_recovery_benign_attack} use $T_w=10$, and $T_c=5$ and the fang distribution. Contrary to the rest of the datasets used, we use $T_c=5$ because CIFAR10 was a much more challenging learning task.


\subsection{MNIST~\cite{lecunmnist}}\label{setup:mnist}
MNIST is a 10-class digit image classification dataset, which contains 70,000 grayscale images of size 28 $\times$ 28. We consider 100 FL clients and divide all data using Dirichlet or FCJ distributions. We use the same CNN as the FEMNIST dataset.

\noindent\textbf{Hyperparameters for re-eval:} For MNIST, we run over 2000 epochs with 100 clients, a learning rate of $0.03$, and a batch size of 32. In the attack setting, 20 clients are malicious. We set $T_w=20$, and $T_c=10$. The FL algorithm used here is FedSGD. The results reported in Figure~\ref{fig:fdr_recovery_benign_attack} use the FCJ distribution.

\subsection{Fashion-MNIST~\cite{xiao2017fashion}}\label{setup:fashion}
Fashion-MNIST is a 10-class image classification dataset with grayscale images of clothing of size 28 $\times$ 28. It contains 70,000 total images. 
We consider 100 FL clients and divide all 70,000 images using Dirichlet or FCJ distributions.
For CIFAR10, MNIST, and FashionMNIST, we divide each client's data in train/test/validation splits in the ratio of $10:1:1$. We combine clients' validation data and use it for validation and hyperparameter tuning and report accuracy on test data. We use the same CNN as the FEMNIST dataset.

\noindent\textbf{Hyperparameters for re-eval:} We run over 2000 epochs with 100 clients, a learning rate of $3\times10^{-3}$~\footnote{We could not achieve the same accuracy reported in~\cite{cao2022fedrecover} using their reported $3\times10^{-4}$ learning rate, hence we use $3\times10^{-3}$.}, and a batch size of $32$. In the attack setting, 20 clients are malicious. We set $T_w=20$, and $T_c=10$. The FL algorithm used here is FedSGD. The results reported in Figure~\ref{fig:fdr_recovery_benign_attack} use the FCJ distribution.

\subsection{StackOverflow~\cite{stackoverflow2019}}\label{setup:stackoverflow}
StackOverflow is a language-modeling dataset that is used for tag prediction and next-word prediction. It consists of 342,477 users who are used as clients, and the training data consists of 135,818,730 examples. We use an RNN with a 96-dimensional embedding and a 10000-word vocabulary. The complete network consists of an input layer followed by an embedding layer, an LSTM layer, and two dense layers.
We use the cross-device setting to obtain the baseline in~\cite{reddi2020adaptive} by using the fedjax~\cite{fedjax2021} framework. The FL algorithm is $FedAdam$, and the training consists of 1500 rounds with 50 clients chosen every round with one local epoch. We keep the batch size to 16, the client optimizer as SGD with a learning rate of $10^{-3}$, and Adam as the server optimizer with a learning rate of $10^{-2}$.
\section{Impact Analysis of Pitfalls}\label{sec:impact}
In this section, we 
analyze the impact of each identified pitfall. We test representative defenses across diverse setups by following the recommendations outlined in \S\ref{sec:pitfalls} and scrutinizing their implications. This systematic exploration is intended to help researchers make informed decisions about the robustness of FL defenses.

\begin{figure}[t]
\centering
\hspace*{-.5cm}
\includegraphics[width=.9\columnwidth]{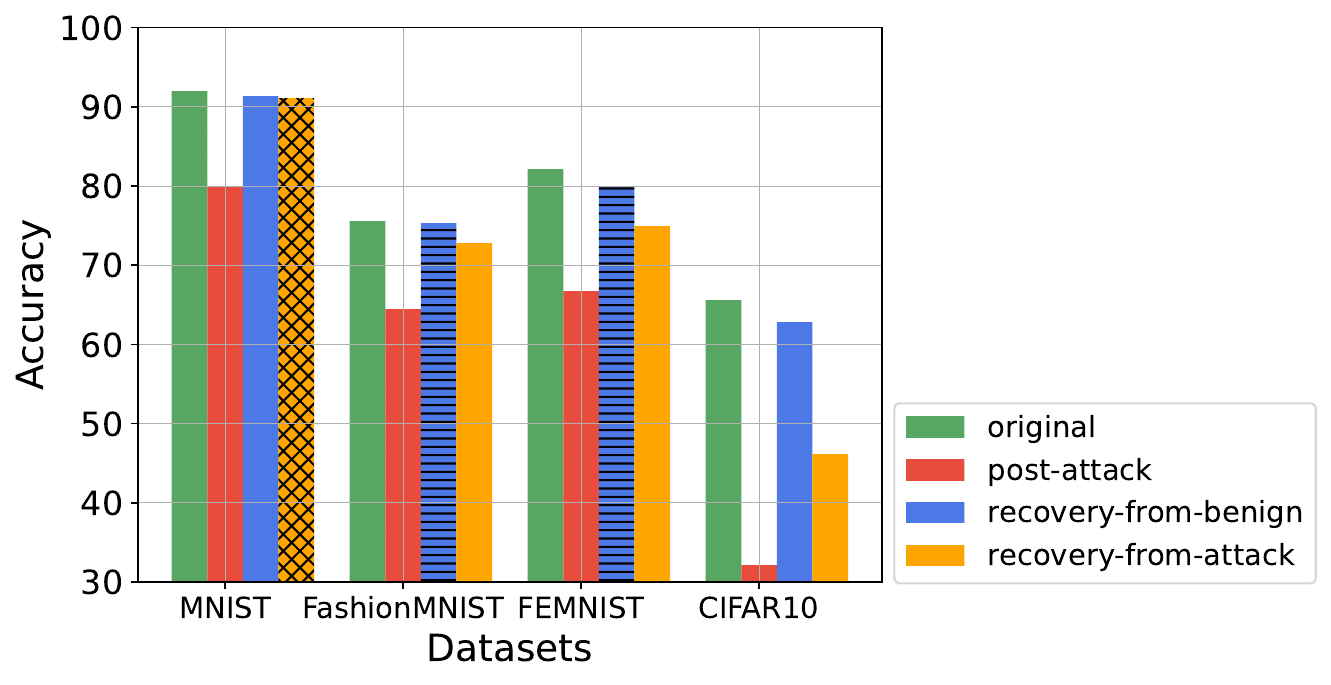}
\caption{Performance of FedRecover on four datasets, both with and without trim attack in the original training.}
\label{fig:fdr_recovery_benign_attack}
\end{figure}
\subsection{Intrinsically Robust Datasets}\label{impact:datasets}
First, we evaluate how the selection of datasets impacts the robustness of our three representative FL defenses: TrMean, FedRecover, and FLDetector.
\subsubsection{TrMean is only robust with MNIST}\label{impact:datasets:trmean}
To assess TrMean's sensitivity to different datasets, we employ FedAvg and FedSGD on MNIST, FashionMNIST, CIFAR10, and FEMNIST, both without attacks and under Stat-Opt~\cite{fang2020local}. Results in Figure~\ref{fig:fl_baseline}a highlight MNIST's intrinsic robustness. Despite a high ($20\%$) amount of malicious clients, MNIST-trained FL model accuracy drops less than $1\%$ in FedAvg, while other datasets experience more substantial declines, peaking at $50\%$ for CIFAR10.

The variation in performance can be attributed to task complexity. 
Baseline accuracies in Figure~\ref{fig:fl_baseline} reflect this complexity, with MNIST having the highest no-attack accuracy and CIFAR10 the lowest.
From this set of experiments, we can conclude that \emph{TrMean is highly robust using MNIST-based evaluations, but not with other datasets as the evaluations of TrMean using other three datasets, FashionMNIST, FEMNIST, and CIFAR10 show.}
\subsubsection{FedRecover works better with simple datasets}\label{impact:datasets:fdr}
We evaluate FedRecover on FashionMNIST, FEMNIST, and CIFAR10, in addition to MNIST, since  MNIST is heavily evaluated in ~\cite{cao2022fedrecover}. We test FedRecover under \emph{recovery-from-benign} and \emph{recovery-from-attack} scenarios. In the former, no attack occurs during original training, while in the latter, Stat-Opt attack is applied during original training but not during recovery (we discuss attack during recovery in \S\ref{impact:attacks:fdr}). Consistent results with~\cite{cao2022fedrecover} on MNIST and FashionMNIST validate our implementation.

Figure~\ref{fig:fdr_recovery_benign_attack} shows that even without attacks, FedRecover does not fully recover for complex datasets like FEMNIST and CIFAR10. In recovery from an attack for MNIST (Fashion-MNIST), FedRecover achieves 91\% (72.7\%) recovery accuracy from a post-attack accuracy of 80\% (64\%), where the original training accuracy is 92\% (75.2\%). For FEMNIST and CIFAR10, differences between the baseline accuracy and recovery accuracies are 11\% and 19\%, respectively, indicating an incomplete recovery in the attack setting. This performance variation stems from dataset complexity, with estimation errors higher for complex tasks such as FEMNIST. The impact of \emph{periodic correction} and \emph{warmup} phase on estimation error and recovery performance is discussed in \S\ref{impact:algorithm}, \S\ref{impact:scalability}, and \S\ref{impact:attacks}.

\begin{table}[]
\caption{Impact of data-level perturbations on FLDetector.}
\scriptsize
\resizebox{.88\columnwidth}{!}{%
\begin{tabular}{|c|c|cc|cc|}
\hline
\multirow{2}{*}{dataset} & \multirow{2}{*}{pertubation} & \multicolumn{2}{c|}{FedSGD} & \multicolumn{2}{c|}{FedAvg} \\ \cline{3-6} 
                         &                & \multicolumn{1}{c|}{FPR} & FNR & \multicolumn{1}{c|}{FPR} & FNR \\ \hline
\multirow{2}{*}{MNIST}   & Noisy-features & \multicolumn{1}{c|}{0}   & 0   & \multicolumn{1}{c|}{0}   & 0   \\ \cline{2-6} 
                         & Noisy-label    & \multicolumn{1}{c|}{0}   & 0   & \multicolumn{1}{c|}{0}   & 0   \\ \hline
\multirow{2}{*}{Fashion} & Noisy-features & \multicolumn{1}{c|}{0}   & 0   & \multicolumn{1}{c|}{0}   & 0   \\ \cline{2-6} 
                         & Noisy-label    & \multicolumn{1}{c|}{0}   & 0   & \multicolumn{1}{c|}{0}   & 0   \\ \hline
\multirow{2}{*}{FEMNIST} & Noisy-features & \multicolumn{1}{c|}{0}   & 1   & \multicolumn{1}{c|}{0}   & 1   \\ \cline{2-6} 
                         & Noisy-label    & \multicolumn{1}{c|}{0}   & 1   & \multicolumn{1}{c|}{0}   & 1   \\ \hline
\end{tabular}%
}
\label{tab:fld_data_distribution}
\end{table}
\subsubsection{FLDetector's performance varies with task complexity}\label{impact:datasets:fld}
\begin{figure}[t]
    \centering
    \begin{subfigure}[b]{1.06\linewidth}
        \hspace{-.6cm}
        \includegraphics[width=\linewidth]{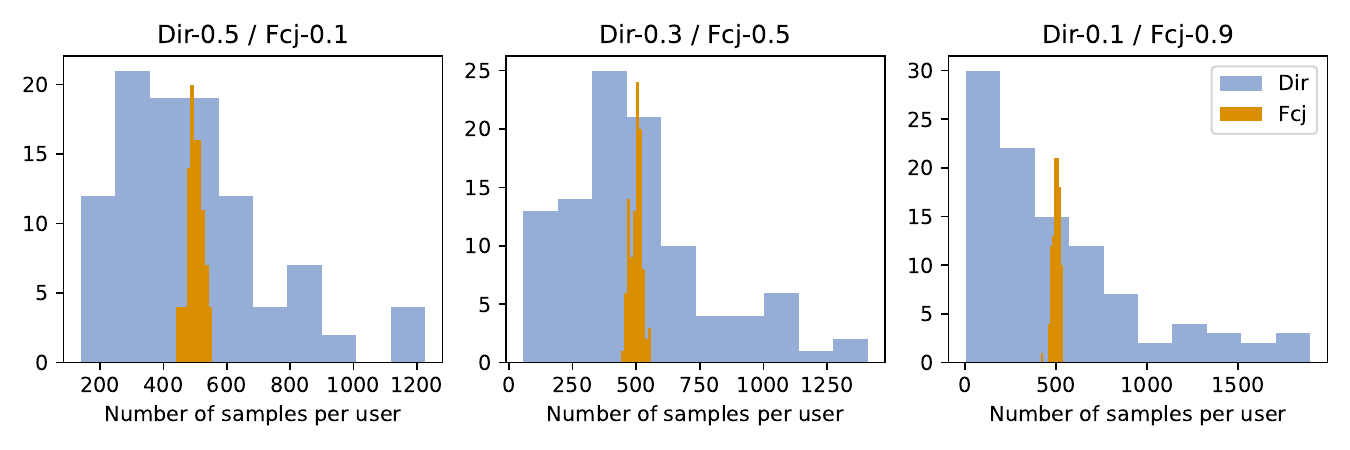}
        \caption{Histogram of number of samples per client.}
        \label{fig:num_samples_per_user}
    \end{subfigure}
    \begin{subfigure}[b]{1.06\linewidth}
        \hspace{-.6cm}
        \includegraphics[width=\linewidth]{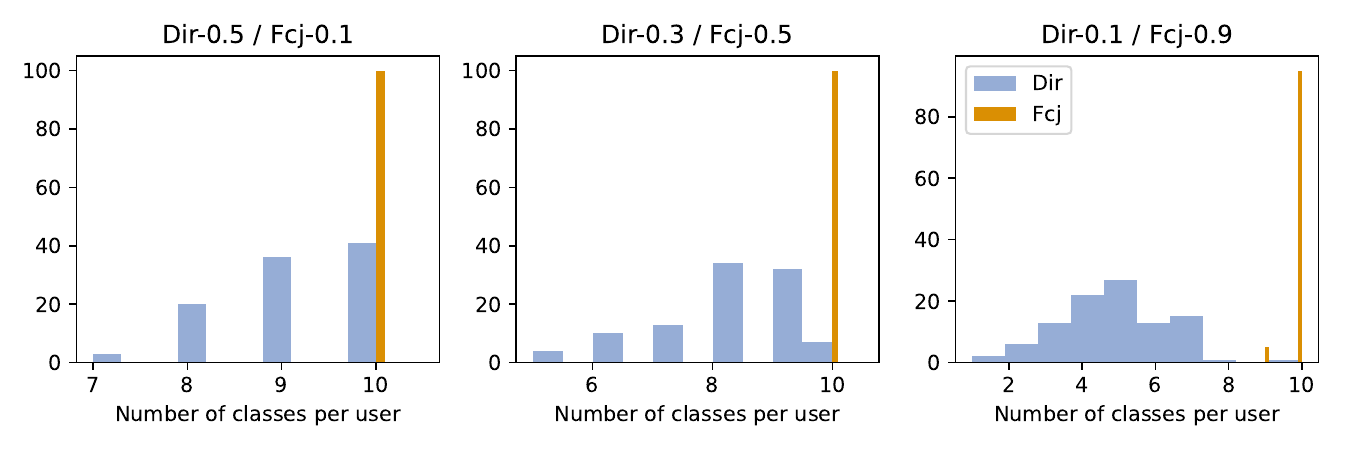}
        \caption{Histogram of number of classes per client.}
        \label{fig:num_classes_per_user}
    \end{subfigure}
    \caption{Comparison of Sample and Class Distribution in FL Datasets: Histograms illustrating (a) the number of samples and (b) the number of classes per client, generated using FCJ and Dir distributions. 
    From left to right, the non-i.i.d. degree of generated datasets increases, reflecting the impact of higher FCJ (or Dir) parameters in generating more (or less) non-i.i.d. datasets. We note that all FCJ client datasets are almost the same size, while Dir client datasets are widely varying. Similarly, with FCJ, all clients have all the classes, while with Dir, the number of classes varies widely.}
    \label{fig:num_samples_classes_per_user}
\end{figure}

We assess FLDetector's performance across varying task complexities using MNIST, FashionMNIST, and FEMNIST. To enhance task complexity, we introduce minor perturbations in features and labels, as we already know the results of these datasets in the unperturbed setting~\cite{zhang2022fldetector}. Table~\ref{tab:fld_data_distribution} shows that MNIST and FashionMNIST datasets remain robust, with FLDetector achieving perfect detection, i.e., zero FNR (False Negative Rate) and FPR (False Positive Rate). However, for FEMNIST, FLDetector shows an FPR of 0 and FNR of 1 across all conditions, indicating that it fails to detect any attacks, classifying all malicious clients as benign and allowing them to remain in the training process.

Performance variations across datasets stem from task complexity, extensively discussed in \S\ref{impact:datasets:trmean} and \S\ref{impact:datasets:fdr}. MNIST and FashionMNIST, less affected by the Stat-Opt attack (Figure~\ref{fig:fl_baseline}), have closely clustered benign updates, making it difficult for malicious updates to be stealthy. This proximity aids FLDetector in distinguishing between malicious and benign updates. Conversely, FEMNIST, a naturally distributed and more heterogeneous dataset than MNIST and FashionMNIST (\S\ref{impact:distribution:statistical_analyses}), results in client updates being further apart. This increased distance enables a malicious update to blend in seamlessly, evading detection by FLDetector.
\subsection{Homogeneous Data Distribution}\label{impact:distribution}
In this section, we first show that the Dir (Dirichlet) distribution is more \emph{real-world} than the FCJ distribution through statistical analyses and visualization. Subsequently, we show the effect of these distributions and their varying levels of heterogeneity on the robustness of TrMean and FedRecover. The original work on FLDetector~\cite{zhang2022fldetector} already analyzed different levels of heterogeneity, so we skip this here.
\subsubsection{Statistical analyses of FCJ and Dir distributions}\label{impact:distribution:statistical_analyses}
We consider a classification task with a total of C classes; we generate client datasets using Dir and FCJ for 100 clients with varying degrees of non-i.i.d. We provide analyses for CIFAR10, but it applies to other datasets. We then plot the following three statistics of the datasets:
\begin{figure}[t]
    \centering
    \begin{subfigure}[b]{\columnwidth}
        \hspace{-.6cm}
        \includegraphics[scale=.6]{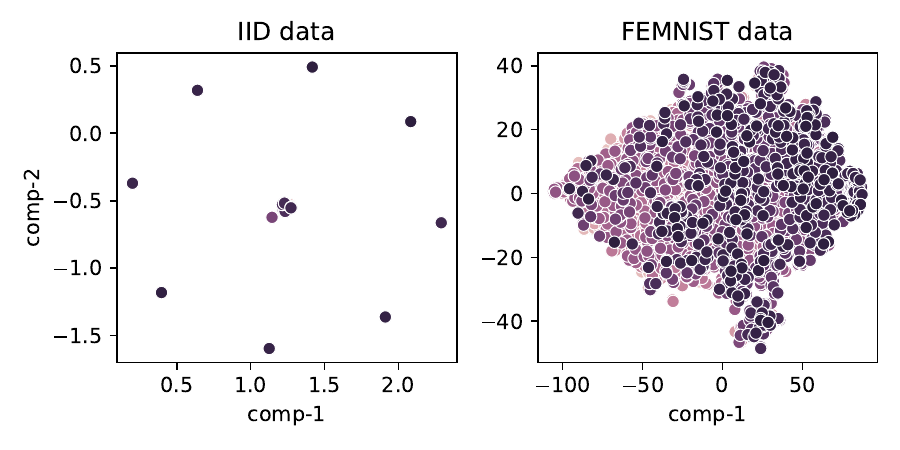}
        \caption{(Left) i.i.d. client datasets, (right) real-world FEMNIST datasets.}
        \label{fig:iid_femnist_tsne}
    \end{subfigure}
    \begin{subfigure}[b]{1\columnwidth}
        \includegraphics[width=\columnwidth]{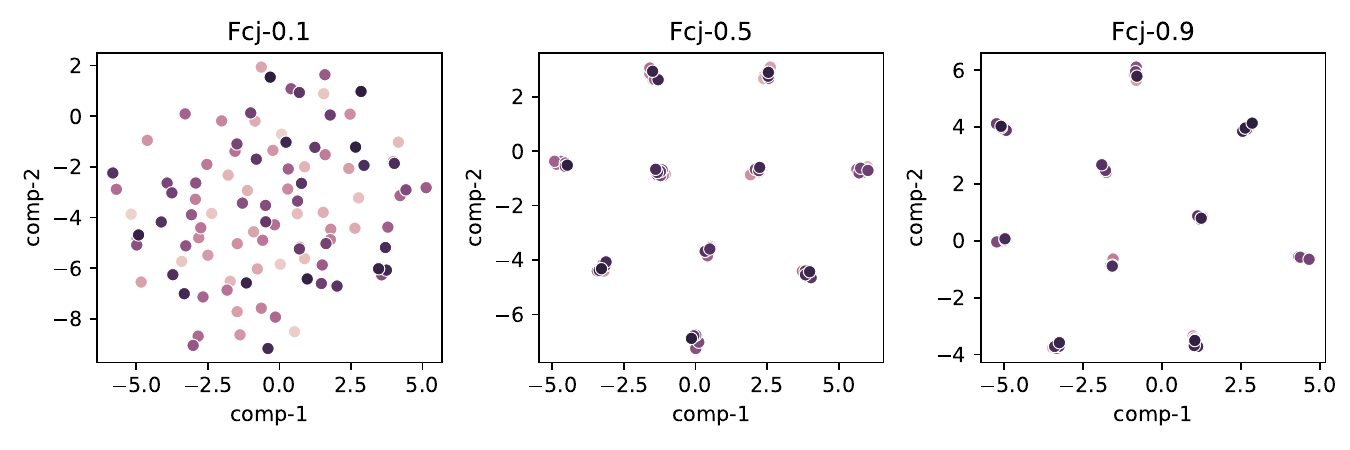}
        \caption{FCJ distributed client datasets.}
        \label{fig:fcj_tsne}
    \end{subfigure}
    \begin{subfigure}[b]{1\columnwidth}
        \includegraphics[width=\columnwidth]{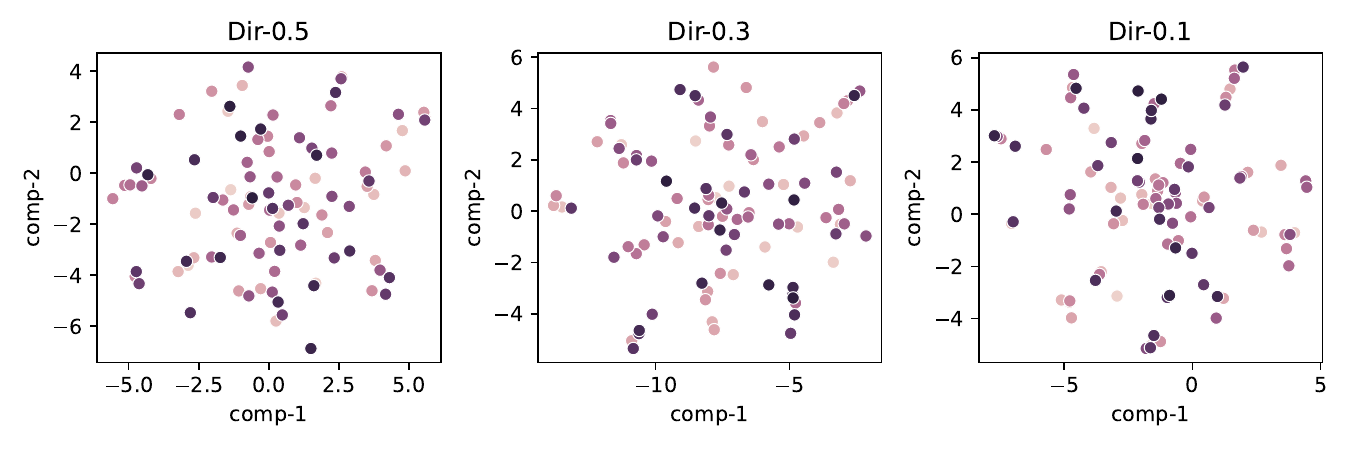}
        \caption{Dir distributed client datasets.}
        \label{fig:dir_tsne}
    \end{subfigure}
    \caption{T-SNE projections of class frequency vectors of client datasets generated using FCJ (b) and Dir (c) distributions. From left to right, non-i.i.d. degree increases.}
    \label{fig:tsne_dir_fcj}
\end{figure}

\noindent \textbf{(Stat-1):} We plot the number of samples per user, which is motivated by the visualizations of real-world FL datasets in Leaf (Figure~\ref{fig:femnist_hist}).
Figure~\ref{fig:num_samples_per_user} shows the results for three degrees of non-i.i.d. For Dir we use $\alpha\in\{0.1, 0.3, 0.5\}$ and for FCJ we use bias $b\in\{0.9, 0.5, 0.1\}$.
We note that Dir produces client datasets with heterogeneous sizes; the histograms are similar to real-world datasets in the Leaf repository (Figure~\ref{fig:femnist_hist}). However, FCJ makes client datasets with almost equal size; note that the FCJ histograms are always concentrated around 500 (the total number of samples in the CIFAR10 dataset / total number of clients).

\noindent \textbf{(Stat-2):} In Figure~\ref{fig:num_classes_per_user}, we show the number of classes each client has when we use Dir or FCJ. We observe that for FCJ, all clients have all the classes except when the bias is 0.9. In the real-world datasets, all clients generally do not have all the classes~\cite{mcmahan2017communication}. Similar to these real-world datasets, the clients have a widely varying number of classes in Dir distribution.

\noindent\textbf{(Stat-3):} For each client, we compute a C-dimensional vector where the $i^{th}$ dimension represents the number of samples from class $i$; here, we use CIFAR10 with 100 clients; hence we get 100 10-dimensional vectors. Then, we plotted T-SNE projections of these 100 vectors; we plotted them for both Dir and FCJ distributed client datasets. Figures~\ref{fig:fcj_tsne} and~\ref{fig:dir_tsne} show the projections for Dir and FCJ, respectively. 
For reference, we also show in Figure~\ref{fig:iid_femnist_tsne} how the T-SNE projections look like for (1) 100 clients with perfectly i.i.d. datasets and (2) the real-world FEMNIST dataset.

For FCJ distribution, we note that, for bias values greater than 0.2 (Figure~\ref{fig:fcj_tsne} center and right), the client datasets form local clusters, i.e., within these clusters, the clients have highly i.i.d. datasets. This is expected because FCJ forms C groups of clients where $i^{th}$ group gets the bias fraction of data from $i^{th}$ class and bias/($C-1$) fraction of data from other classes. This data is then randomly assigned to clients within the $i^{th}$ group, which makes these clients' datasets i.i.d. On the other hand, globally,  
these clusters form potentially non-i.i.d. structures. However, for a bias of 0.1, we observe almost i.i.d. datasets as expected; Figure~\ref{fig:iid_femnist_tsne}-left shows a perfectly i.i.d. dataset. To summarize, \emph{although both are globally non-i.i.d., we have shown that the FCJ distribution is locally-i.i.d., while Dir is locally non-i.i.d.}. Next, we study their impact.
\subsubsection{TrMean is more robust with lower heterogeneity}\label{impact:distribution:trmean}
\begin{figure}[t]
    \centering
    \begin{subfigure}[b]{0.48\columnwidth}
        \includegraphics[width=\columnwidth]{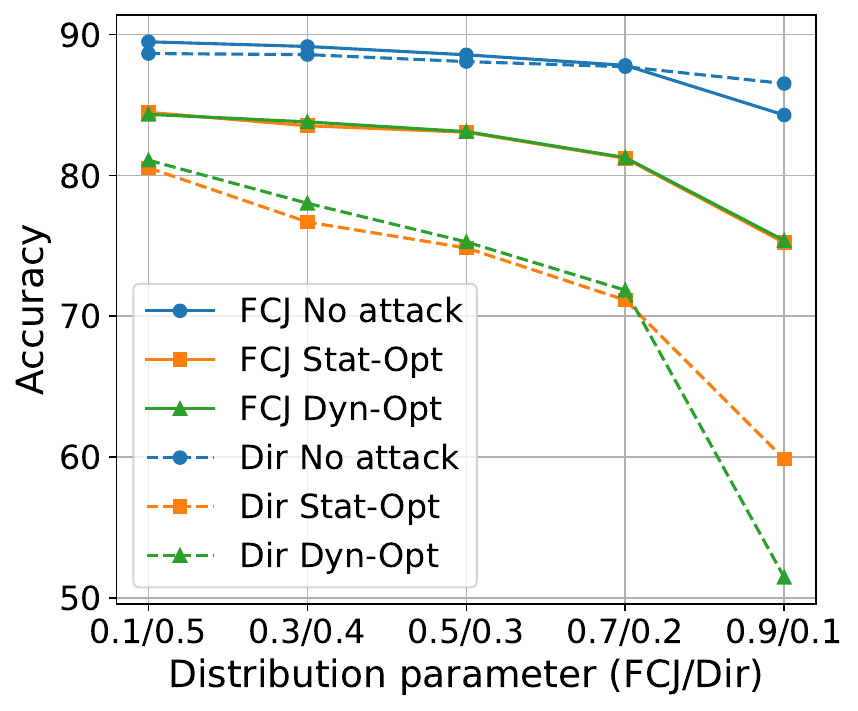}
        \caption{TrMean}
        \label{fig:trmean_distributions_fmnist}
    \end{subfigure}
    \begin{subfigure}[b]{0.48\columnwidth}
        \includegraphics[width=\columnwidth]{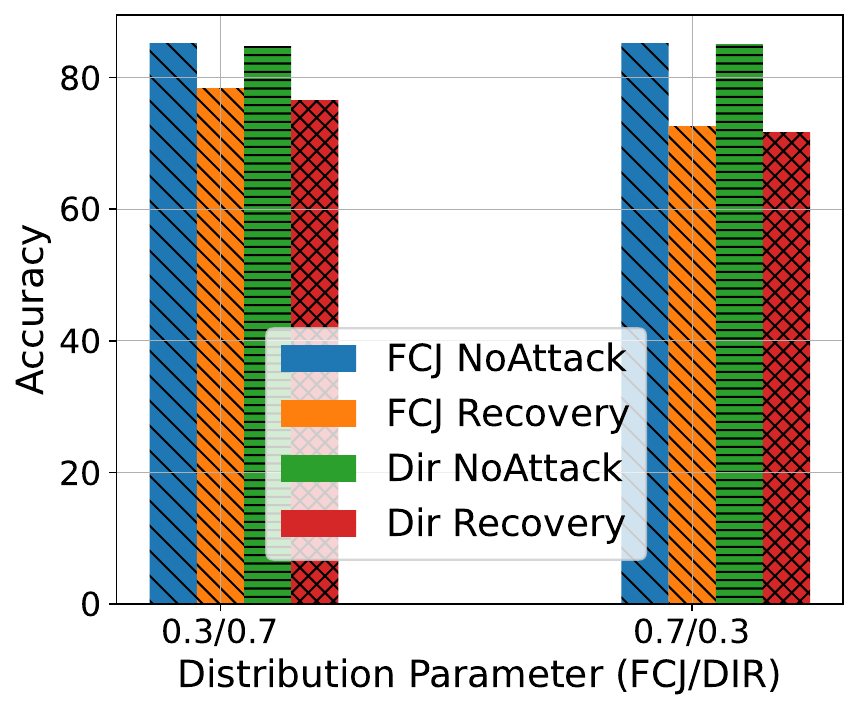}
        \caption{FedRecover}
        \label{fig:fedrecover_distributions_fmnist}
    \end{subfigure}
    \caption{The effect of varying heterogeneity levels for FCJ and Dir distributions on the FashionMNIST dataset.}
    \label{fig:distributions_trmean_fedrecover}
\end{figure}

We demonstrate the impact of FCJ and Dir on TrMean's robustness for FashionMNIST\footnote{We observe similar trends for other datasets, but for brevity, we only include FashionMNIST here.} in Figure~\ref{fig:trmean_distributions_fmnist}. In the no-attack setting, FCJ (DIR) shows little difference, with accuracy going from $89\%$ ($88\%$) at $0.1$ ($0.5$) bias to $84\%$ ($86\%$) at $0.9$ ($0.1$) bias. However, the distinction emerges in the attack setting where FCJ is more robust. For the Stat-Opt attack, FCJ (DIR) accuracy decreases from $84\%$ ($80\%$) at $0.1$ ($0.5$) bias to $75\%$ ($59\%$) at $0.9$ ($0.1$) bias. A similar trend is observed for the Dyn-Opt attack. The performance change is rooted in the nature of the distributions (\S\ref{impact:distribution:statistical_analyses}). FCJ is locally i.i.d. but globally non-i.i.d., while Dir is non-i.i.d. both locally and globally. Due to Dir's greater heterogeneity, benign updates are more dispersed, making it easier for a malicious update to hide. Consequently, TrMean struggles to detect malicious updates, leading to lower global model performance.


\subsubsection{FedRecover performs better under lower heterogeneity}\label{impact:distributions:fdr}
We show the impact of FCJ and Dir distributions on FedRecover for the FashionMNIST dataset 
in Figure~\ref{fig:fedrecover_distributions_fmnist}.
We observe that the performance of FedRecover lowers when we increase the level of heterogeneity and that FCJ performs slightly better than Dir. Specifically, the difference between no-attack accuracies for FCJ at bias 0.3 and 0.7 is 7\% and 12.6\%, respectively, whereas, for Dir, that difference is slightly higher; 8.2\% and 13.3\% corresponding to non-i.i.d. parameters of 0.7 and 0.3 respectively\footnote{For FCJ, a higher value of the i.i.d. parameter means higher heterogeneity, but it is the opposite case for Dir.}.

The reason for FedRecover's performance reduction under high heterogeneity is that a higher heterogeneity means that client updates are far apart and drift away from the global model. Since \emph{the update estimation in FedRecover uses knowledge of the current and past global models, the HVP calculation step (\S\ref{background:fdr}) incurs significant estimation errors if the local model drifts away from the global model.}
\subsection{Slow-converging Algorithms}\label{impact:algorithm}
Here, we will study the impact of the choice of the two widely used FL algorithms, FedSGD and FedAvg, on the robustness of TrMean and FedRecover. We combine our analysis of algorithms with the choice of attacks for FLDetector in \S\ref{impact:attacks:fld}; therefore, we do not discuss it in this section.
\subsubsection{Fast algorithms make TrMean more robust}\label{impact:algorithm:trmean}
In Figure~\ref{fig:fl_baseline}, we show the accuracy of FedSGD and FedAvg with TrMean in benign and Stat-Opt attack scenarios. For FedSGD, we align our implementation with recent defenses~\cite{cao2022fedrecover,fang2020local}. In contrast, FedAvg is optimized for greater accuracy with only $5\%$ to $20\%$ of FedSGD's communication (the number of rounds).

Under both benign and adversarial conditions, FedAvg greatly surpasses FedSGD in performance, convergence, and communication for all datasets, as shown in Figure~\ref{fig:fl_baseline}. 
We can conclude here that in adversarial situations, \emph{FedAvg proves more resilient to untargeted poisoning due to its rapid convergence}, giving adversaries minimal time for poisoning.
\subsubsection{Fast algorithms lower estimation errors in FedRecover}\label{impact:algorithm:fdr}
We examine FedRecover's performance under different algorithms, FedSGD and FedAvg. The original study~\cite{cao2022fedrecover} applies the Stat-Opt attack to MNIST with FedSGD over 2000 rounds, reducing accuracy from 96\% to 81\%. By employing FedAvg with carefully chosen hyperparameters (Appendix~\ref{sec:setup}), as shown in Figure~\ref{fig:baseline_mnist}, we achieve 98\% accuracy in 50 rounds, with the Stat-Opt attack lowering it to only 96\%. This allows for perfect recovery, as depicted in Figure~\ref{fig:acc_vs_comm_mnist_fashion}, where both MNIST and FashionMNIST (no-attack accuracy: 87\%) show perfect recovery under the fast-converging FedAvg. This perfect recovery holds even with variations in the periodic correction periodicity $T_c$.
FedRecover excels in this scenario because \emph{FedAvg's rapid convergence (\S\ref{background:FL}) provides a high starting accuracy during FedRecover's \emph{warmup phase}.} Throughout periodic correction and abnormality fixing phases, \emph{client updates computed over multiple local epochs (in contrast to FedSGD's single local epoch) lead to lower estimation errors, ensuring perfect recovery.}
\subsection{Limited FL Settings}\label{impact:scalability}
To show the effect of \emph{scale} on FL defenses, we first assess the impact of a scale-constrained threat model on TrMean. We then show why FedRecover and FLDetector are incompatible with cross-device settings and discuss the practicality of storage, computation, and communication for FedRecover (or, any mechanism requiring stored historical information).

\begin{figure}
    \centering
    \begin{subfigure}[b]{0.48\columnwidth}
        \includegraphics[width=\columnwidth]{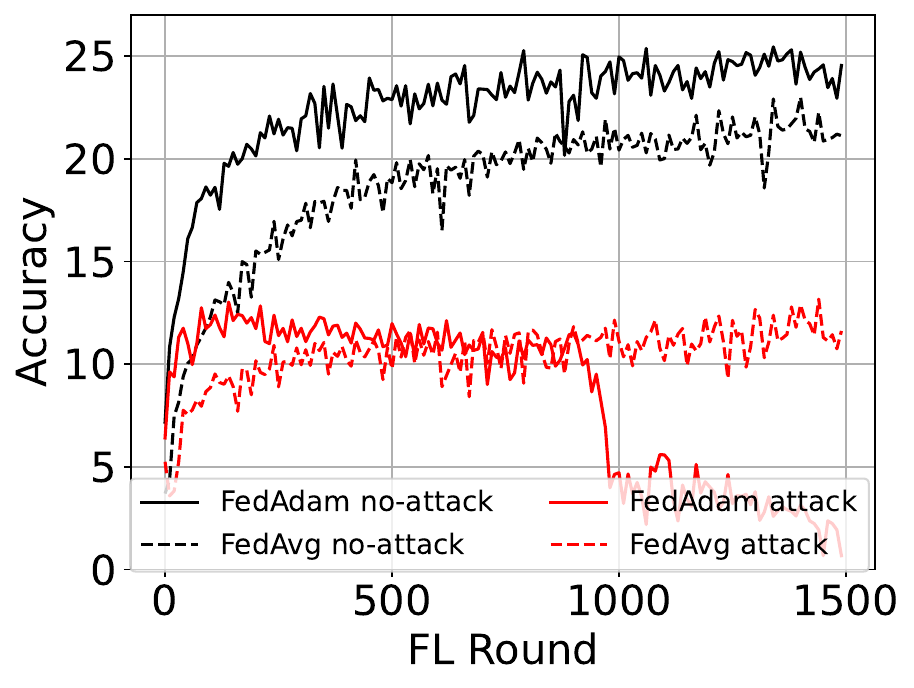}
        \caption{Trim Attack}
        \label{fig:stackoverflow_trim}
    \end{subfigure}
    \begin{subfigure}[b]{0.48\columnwidth}
        \includegraphics[width=\columnwidth]{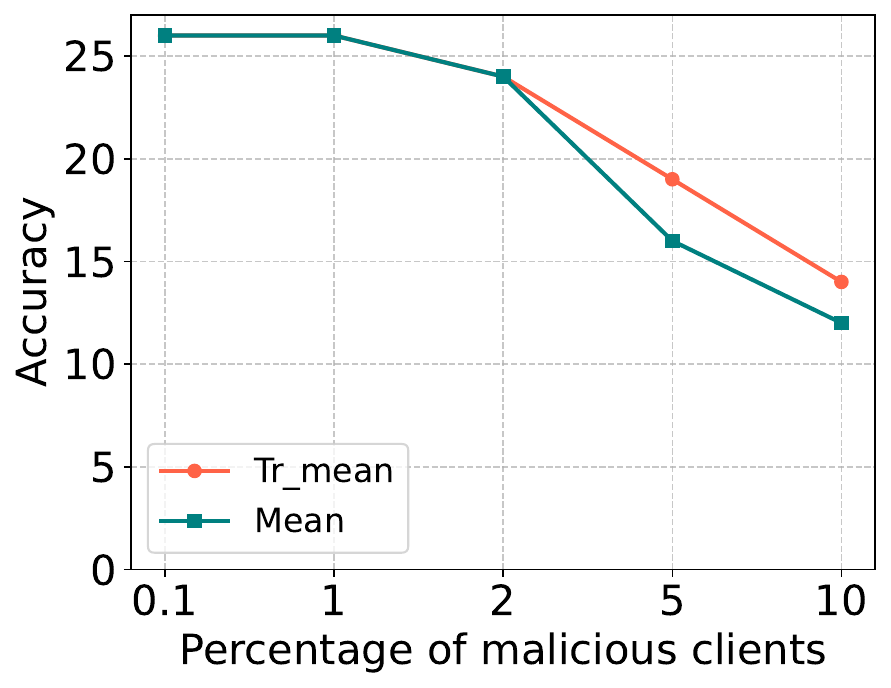}
        \caption{Varying \% malicious}
        \label{fig:stackoverflow_percentage_malicious}
    \end{subfigure}
    \caption{Stackoverflow in practical settings.}
    \label{fig:stackoverflow}
\end{figure}
\subsubsection{Mean AGR is robust on a large scale}\label{impact:scalability:stackoverflow}
In this section, we showcase the robustness of the mean AGR, typically not robust in cross-silo settings~\cite{yin2018byzantine}, within a large-scale, cross-device environment. \emph{Given the growing popularity of language models~\cite{touvron2023llama, brown2020language}, although underrepresented in our defense survey (Figure~\ref{fig:defense_survey}a), we leverage the StackOverflow~\cite{stackoverflow2019} dataset for large-scale evaluation} (setup details in Appendix~\ref{sec:appendix}). The no-attack baselines in Figure~\ref{fig:stackoverflow_trim} are obtained following the settings in~\cite{reddi2020adaptive}. We can see that with the standard amount of 20\% malicious clients, the Stat-Opt attack significantly impacts StackOverflow in cross-device.

The StackOverflow FL setting has about 300,000 clients and $20\%$ amounts to 60,000 malicious clients. Accessing and modifying such a vast number of devices is considered impractical, factoring in operational and financial costs~\cite{shejwalkar2022back}. Figure~\ref{fig:stackoverflow_percentage_malicious} shows a decline in attack performance as the percentage of malicious clients decreases. Notably, \emph{\textbf{with less than 5\% malicious clients, the Mean AGR remains unaffected}} by the Stat-Opt attack.
\emph{It is important to highlight the difference between our findings and~\cite{shejwalkar2022back}} where the attack impact with Mean AGR is $>0\%$ even for $0.01\%$ of malicious clients for FEMNIST, CIFAR10, and Purchase, while in our Stack Overflow case, the attack impact is $0\%$ for $<2\%$ malicious clients. With our results, \emph{\textbf{we strongly emphasize using datasets especially designed for FL and evaluating defenses under constraints imposed by scaling up the system in addition to small-scale experimentation.}} We do not dismiss the significance of small-scale experiments. Cross-silo
FL is indeed widely used. Instead, we emphasize the critical need for evaluating FL defenses within scaled-up settings.
\subsubsection{Resource overheads for FedRecover}\label{impact:scalability:fdr_fld}
Here, we first explain the reasons behind our conviction that FedRecover and FLDetector are incompatible in the cross-device setting. Consequently, 
we do not evaluate these two defenses under the cross-device setting. Nevertheless, in this section, we comment on some of the \emph{practical} aspects of FedRecover, such as computation, communication, and storage costs, to assess the feasibility of scaling up such systems.

\noindent\textbf{Compatibility of Fedrecover and FLDetector with the cross-device FL:}
FedRecover's reliance on historical information from clients' past updates is hindered in cross-device, where clients participate in a few rounds, limiting available historical updates. Similarly, FLDetector faces challenges in the cross-device setting due to a lack of consistent historical information. Therefore, \emph{we find FedRecover and FLDetector incompatible with the cross-device setting}.

\begin{figure}[t]
    \centering
    \begin{subfigure}[b]{0.48\columnwidth}
        \includegraphics[width=\columnwidth]{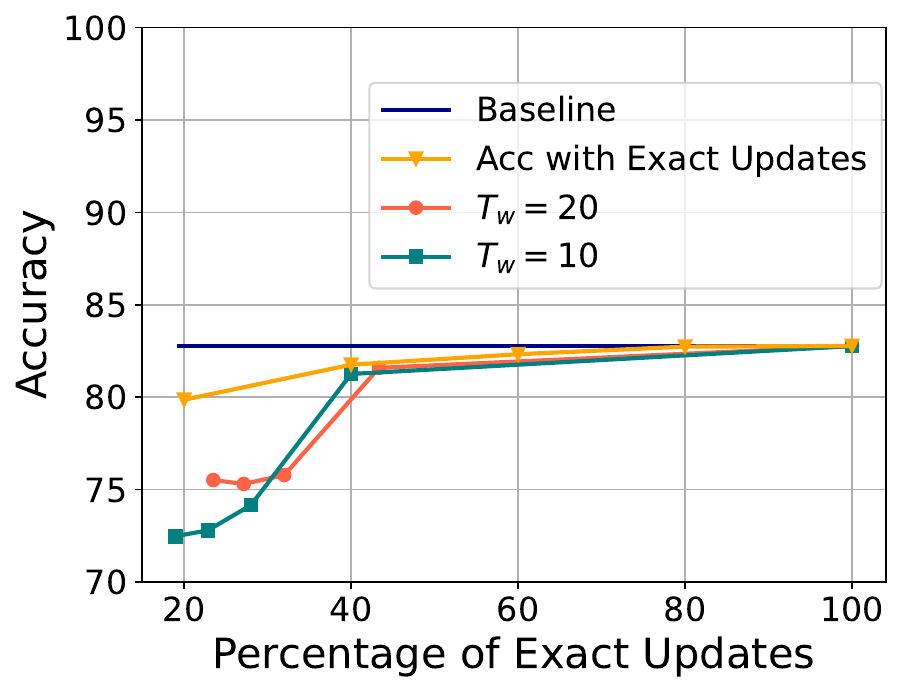}
        \caption{FEMNIST}
        \label{fig:acc_vs_comm_femnist}
    \end{subfigure}
    \begin{subfigure}[b]{0.48\columnwidth}
        \includegraphics[width=\columnwidth]{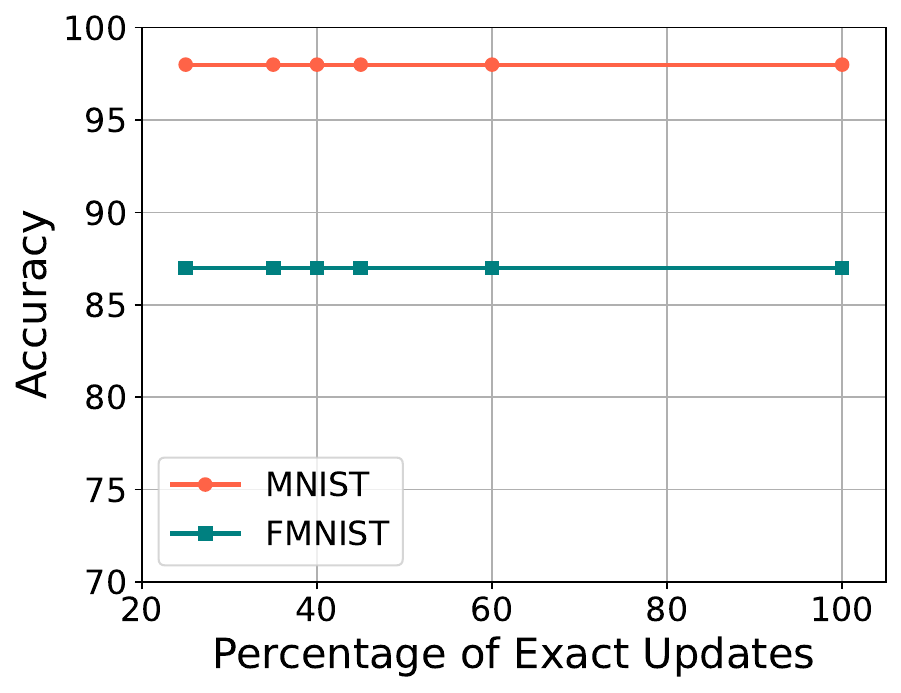}
        \caption{MNIST and Fashion}
        \label{fig:acc_vs_comm_mnist_fashion}
    \end{subfigure}
    \caption{Communication-accuracy tradeoff for FedRecover. Despite the presence of four lines, their overlap is discernible as we can achieve baseline accuracy with exact updates.}
    \label{fig:acc_vs_comm}
\end{figure}

\noindent\textbf{Communication-accuracy tradeoff for FedRecover:}
In Figure~\ref{fig:acc_vs_comm_femnist}, we show the tradeoff between the recovery accuracy and communication of FedRecover.
The recovery accuracy increases as it relies more on exact updates (locally computed updates instead of estimated ones) from the clients, where the \emph{minimum number of exact updates} is $T_w$ + $\frac{T_{total} - T_w - T_f}{T_c}$ + $T_{f}$. Interestingly, using the same amount of exact updates (effectively the same amount of communication as the server would communicate with clients for the same number of rounds) gives us more accuracy by training from scratch, i.e., not using FedRecover but restarting training with benign clients only. For instance, with only $20\%$ exact updates and $T_w=20$, FedRecover achieves $\approx76\%$ accuracy while training from scratch with an equivalent 40 rounds achieves $\approx80\%$.

In Figure~\ref{fig:acc_vs_comm_mnist_fashion}, we can see that with a small percentage of exact updates, FedRecover completely recovers for simpler datasets like MNIST and FashionMNIST with FedAvg and is unaffected by the variation in periodicity $T_c$. With the same amount of exact updates, we can achieve the same results without FedRecover as well. This is because a fast algorithm gives less time for the adversary to attack and a higher starting point for FedRecover to recover, leading to lower estimation errors (\S\ref{impact:algorithm:fdr}).

Based on these observations, we conclude that with fast converging algorithms and proper settings, we might not need to use FedRecover, especially when it comes with an additional computational and storage cost. In our experiments, for the slow baseline that uses FedSGD over 1000 epochs for MNIST, we require \textbf{\emph{$\approx200GB$}} of storage for saving the client model updates every round. By extension, this applies to any defense that requires knowledge of past updates. This cost would significantly increase with the number of clients and by using larger models for more complex tasks.
\subsection{Naive Attacks}\label{impact:attacks}
This component is critical in our research, given the prevalence of simple attacks highlighted in \S\ref{pitfalls:5}. We demonstrate TrMean's vulnerability to powerful model poisoning attacks such as Stat-Opt and Dyn-Opt. We also test our adaptive attack on FLDetector, revealing high rates of \emph{imperfect client detection} and showcasing its impact on FedRecover, which relies on FLDetector to identify malicious clients before recovery.
\subsubsection{Choice of attacks greatly impact TrMean's performance}\label{impact:attacks:trmean}
We conduct Stat-Opt~\cite{fang2020local} and Dyn-Opt~\cite{shejwalkar2021manipulating} attacks on FashionMNIST, varying the heterogeneity for both FCJ and Dir distributions. The results are depicted in Figure~\ref{fig:trmean_distributions_fmnist}.
Generally, Dyn-Opt is stronger than Stat-Opt, particularly noticeable at higher heterogeneity levels. For instance, at Dir bias level $0.1$, the accuracy drops to $51\%$ for Dyn-Opt compared to $59\%$ for Stat-Opt. Dyn-Opt's strength lies in finding optimal perturbations tailored for the dataset at every FL round, making it more potent compared to the static nature of perturbation in Stat-Opt, as discussed in \S\ref{background:attacks_study}.
\begin{figure}[t]
    \centering
    \begin{subfigure}[b]{0.44\columnwidth}
        \includegraphics[width=\columnwidth]{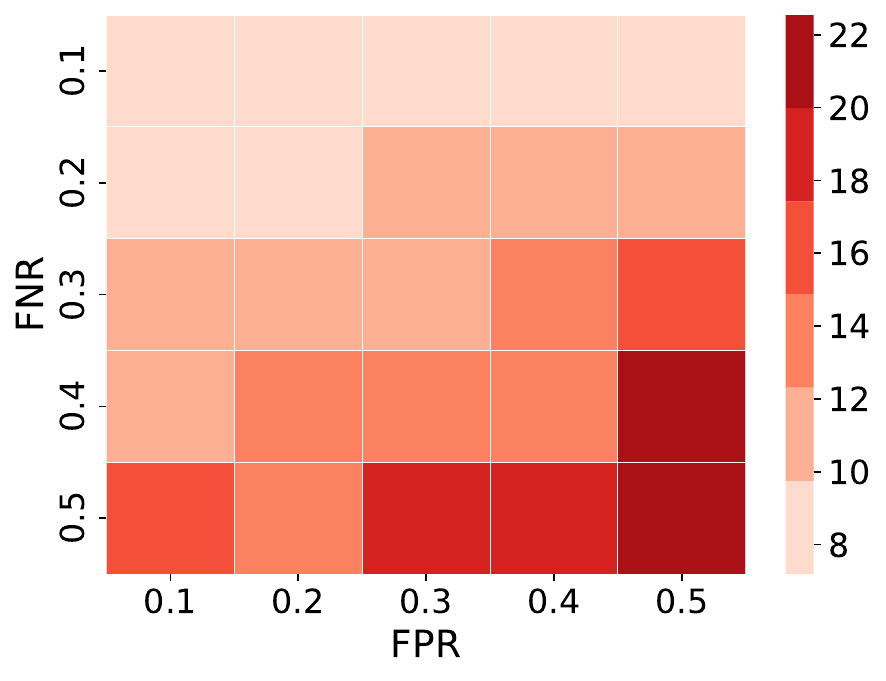}
        \caption{$T_w=10$}
        \label{fig:fedrec_fnr_fpr_twarm20}
    \end{subfigure}
    \hspace{0.5mm}
    \begin{subfigure}[b]{0.44\columnwidth}
        \includegraphics[width=\columnwidth]{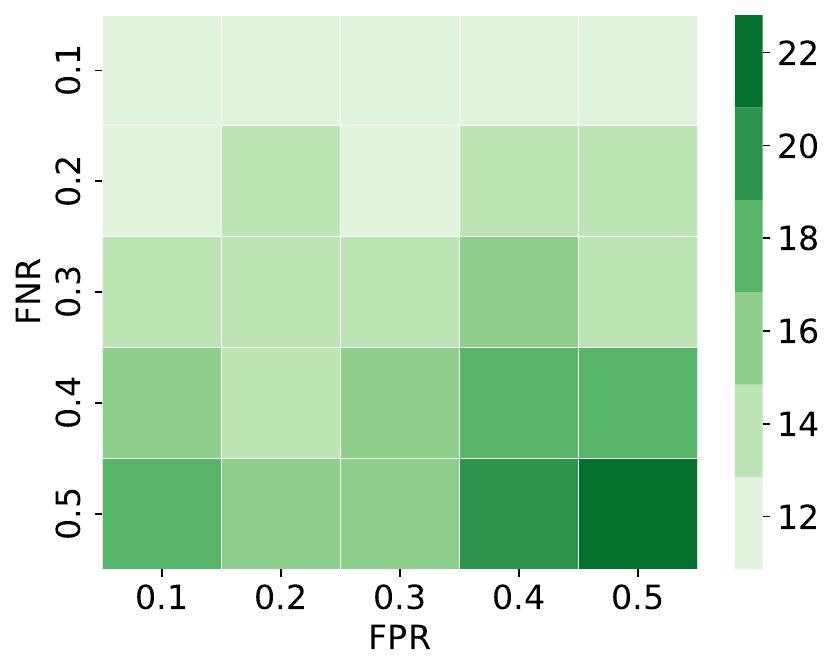}
        \caption{$T_w=20$}
        \label{fig:fedrec_fnr_fpr_twarm10}
    \end{subfigure}
    \caption{Performance of FedRecover, represented by the difference between no-attack and post-recovery accuracies, under non-zero FPRs and FNRs caused by adaptive attacks.}
    \label{fig:fedrec_fnr_fpr}
\end{figure}
\subsubsection{Overcoming FLDetector with our adaptive attack}\label{impact:attacks:fld}
We introduce a novel attack to assess FLDetector's resilience against \emph{adaptive attacks}. Malicious clients craft updates using our attack and evade detection by FLDetector, resulting in non-zero FNRs and FPRs.

Our attack adds a carefully crafted perturbation vector to client model updates, so they are close enough to the estimated model updates, thereby bypassing FLDetector's detection. Here we describe our attack formulation in detail.


\noindent\textbf{\em Attack Formulation:}
FLDetector computes the estimated update (\S\ref{background:fld}) for client $k$ as:
\begin{equation}\label{eqn:fld_update}
    \hat{\nabla}_{t}^{k} = \nabla_{t-1}^{k} + \hat{H}^{t}\cdot(\theta_{t} - \theta_{t-1})
\end{equation}
In this attack, we introduce a perturbation vector, $\mathcal{P}$, that modifies the updates sent from malicious clients. A malicious client computes its update so that its final update is computed as the sum of its previous update, the $HVP$ or $Hessian\ Vector\ Product$ ($\hat{H}^{t}\cdot(\theta_{t} - \theta_{t-1})$), and the perturbation vector. This can be written as:
\begin{equation}\label{eqn:fld_mal_update}
    \hat{\nabla}_{t}^{k} = \nabla_{t-1}^{k} + \hat{H}^{t}\cdot(\theta_{t} - \theta_{t-1}) + \mathcal{P}
\end{equation}

The server estimates an update by adding the $HVP$ to the last round's exact update and compares it with the actual update in that round (\S\ref{background:fdr}). The malicious update, therefore, deviates from the estimated update by $\mathcal{P}$. We proceed to detail the calculation of this perturbation vector $\mathcal{P}$. To calculate $\mathcal{P}$, we first calculate a \emph{good distance range}, $\mathcal{R}$, that is a safe perturbation distance for the perturbation vector by taking the norm between the old and new client updates. The good distance range, $\mathcal{R}_{t}^{k}$, for client $k$, at round $t$ is given by:

\begin{equation}\label{eqn:fld_good_distance}
    \mathcal{R}_{t}^{k} = ||\hat{\nabla}_{t}^{k} - \nabla_{t}^{k}||
\end{equation}

Here, $\hat{\nabla}_{t}^{k}$ is the estimated update for client $k$ at round $t$, and $\nabla_{t}^{k}$ is the actual update for client $k$ at round $t$.
The deviation, $\mathcal{D}$, for the perturbation vector is calculated by following deviation strategies in~\cite{shejwalkar2021manipulating}. It can either be \textit{unit vector}, \textit{sign}, or \textit{std}. Finally, $\mathcal{P}$ is computed by taking the average of all the \textit{good distance ranges}, $\mathcal{R}$, and directing it in the direction of the deviation, $\mathcal{D}$:

\begin{equation}\label{eqn:fld_perturbation}
    \mathcal{P} = \frac{\mathcal{D}}{||\mathcal{D}||}\cdot\frac{1}{N}\sum_{k=1}^{N} \mathcal{R}_{t}^{k}
\end{equation}

The impact of our adaptive attack on the performance of FLDetector is shown in Table~\ref{tab:fld_adaptive_attacks} and across all cases (different combinations of dataset, FL algorithm, and percentage of compromised clients), except one, we find that the FNR is non-zero, which means that malicious clients have not been detected, and they continue to be part of the training process. Since the attack is designed to craft malicious updates that are statistically close to the benign ones, it leads to a non-zero FPR as well, which means that many benign clients are falsely detected as malicious and are removed from the training process.
Since \emph{we achieve a 0 FPR and a 1 FNR for FEMNIST, this means that the attack never gets detected, no benign or malicious client is removed, and all malicious clients seem benign to FLDetector on the server.} We discuss the impact of non-zero FNR and FPR on further training and recovery in \S\ref{impact:attacks:fdr}.

\begin{table}
\caption{Impact of our adaptive attack on FLDetector. Here $\% m$ represents the percentage of malicious clients.}
\centering
\setlength{\tabcolsep}{3pt} 
\begin{tabular}{|c|l|cc|cc|cc|cc|}
\hline
\multicolumn{1}{|l|}{\multirow{2}{*}{\begin{tabular}[c]{@{}l@{}}\textbf{\%}\\ \textbf{m}\end{tabular}}} & \multirow{2}{*}{\textbf{Baseline}} & \multicolumn{2}{l|}{\textbf{MNIST}} & \multicolumn{2}{l|}{\textbf{Fashion}} & \multicolumn{2}{l|}{\textbf{CIFAR10}} & \multicolumn{2}{l|}{\textbf{FEMNIST}} \\ \cline{3-10} 
\multicolumn{1}{|l|}{} &  & \multicolumn{1}{l|}{\textit{FPR}} & \multicolumn{1}{l|}{\textit{FNR}} & \multicolumn{1}{l|}{\textit{FPR}} & \multicolumn{1}{l|}{\textit{FNR}} & \multicolumn{1}{l|}{\textit{FPR}} & \multicolumn{1}{l|}{\textit{FNR}} & \multicolumn{1}{l|}{\textit{FPR}} & \multicolumn{1}{l|}{\textit{FNR}} \\ \hline
 \hline
\multirow{2}{*}{5} & FedSGD & \multicolumn{1}{c|}{0.4} & 1 & \multicolumn{1}{c|}{0.52} & 0.4 & \multicolumn{1}{c|}{0.21} & 0 & \multicolumn{1}{c|}{0} & 1 \\ \cline{2-10} 
 & FedAvg & \multicolumn{1}{c|}{0.02} & 1 & \multicolumn{1}{c|}{0} & 1 & \multicolumn{1}{c|}{0.13} & 1 & \multicolumn{1}{c|}{0} & 1 \\ \hline
  \hline
\multirow{2}{*}{10} & FedSGD & \multicolumn{1}{c|}{0.39} & 1 & \multicolumn{1}{c|}{0.56} & 0.6 & \multicolumn{1}{c|}{0.37} & 0 & \multicolumn{1}{c|}{0} & 1 \\ 
\cline{2-10} 
 & FedAvg & \multicolumn{1}{c|}{0.02} & 1 & \multicolumn{1}{c|}{0.02} & 1 & \multicolumn{1}{c|}{0.23} & 0.67 & \multicolumn{1}{c|}{0} & 1 \\ \hline
  \hline
\multirow{2}{*}{15} & FedSGD & \multicolumn{1}{c|}{0.48} & 0.8 & \multicolumn{1}{c|}{0.54} & 0.3 & \multicolumn{1}{c|}{0.35} & 0.35 & \multicolumn{1}{c|}{0} & 1 \\ \cline{2-10} 
 & FedAvg & \multicolumn{1}{c|}{0.02} & 1 & \multicolumn{1}{c|}{0.05} & 1 & \multicolumn{1}{c|}{0.18} & 0.67 & \multicolumn{1}{c|}{0} & 1 \\ \hline
 \hline
\multirow{2}{*}{20} & FedSGD & \multicolumn{1}{c|}{0.45} & 1 & \multicolumn{1}{c|}{0.62} & 0.3 & \multicolumn{1}{c|}{0.35} & 0.33 & \multicolumn{1}{c|}{0} & 1 \\ \cline{2-10} 
 & FedAvg & \multicolumn{1}{c|}{0.01} & 1 & \multicolumn{1}{c|}{0.06} & 1 & \multicolumn{1}{c|}{0.17} & 0.67 & \multicolumn{1}{c|}{0} & 1 \\ \hline
\end{tabular}
\label{tab:fld_adaptive_attacks}
\end{table}
\subsubsection{Imperfect detection leads to lower recovery performance}\label{impact:attacks:fdr}
Our adaptive attack results in imperfect client detection (\S\ref{impact:attacks:fld}), allowing escaped clients into the recovery process. The escaped clients are denoted by the FNR, while benign clients, incorrectly classified as malicious and subsequently removed from training, are represented by the FPR.

Figure~\ref{fig:fedrec_fnr_fpr} shows that non-zero FNRs and FPRs challenge FedRecover's ability to reach the no-attack accuracy of $82\%$. For instance, at $FNR=FPR=0.3$ for $T_w=10$, the recovery accuracy is only $67\%$. This is because malicious updates in the recovery process cause higher estimation errors and deviate the recovery model from the benign one. A marginal improvement is observed with more warmup rounds, achieving a post-recovery accuracy of $70\%$ for $T_w=20$, as increased warmup rounds provide a higher starting point for the model, resulting in lower overall estimation errors.
\subsection{Unfair Metrics}\label{impact:evaluation}
A key feature of FL is heterogeneity (\S\ref{impact:distribution:statistical_analyses}), which makes testing and reporting individual client's performance essential. Unfortunately, almost all of the surveyed works only report the global model accuracy (\S\ref{pitfalls:6}). To show that this is unfair, we show how personalized performance differs from that of the global model using TrMean and FedRecover under benign and adversarial settings.
\subsubsection{Different clients have different levels of robustness under TrMean}\label{impact:evaluation:trmean}

\begin{figure}[t]
    \centering
    \begin{subfigure}[b]{0.43\columnwidth}
        \includegraphics[width=\columnwidth]{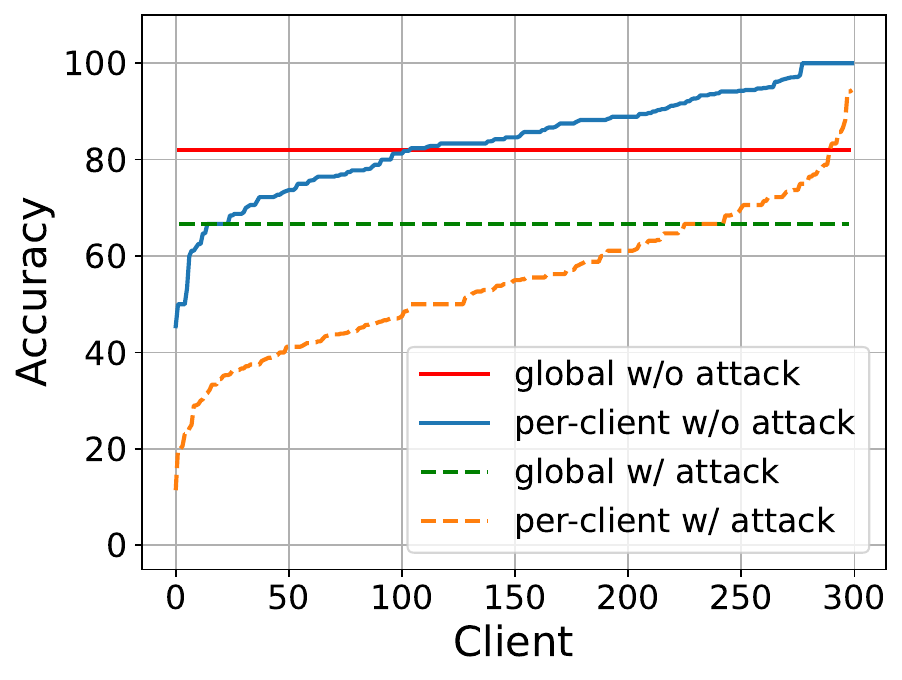}
        \caption{TrMean}
        \label{fig:trmean_per_client_acc}
    \end{subfigure}
    \begin{subfigure}[b]{0.43\columnwidth}
        \includegraphics[width=\columnwidth]{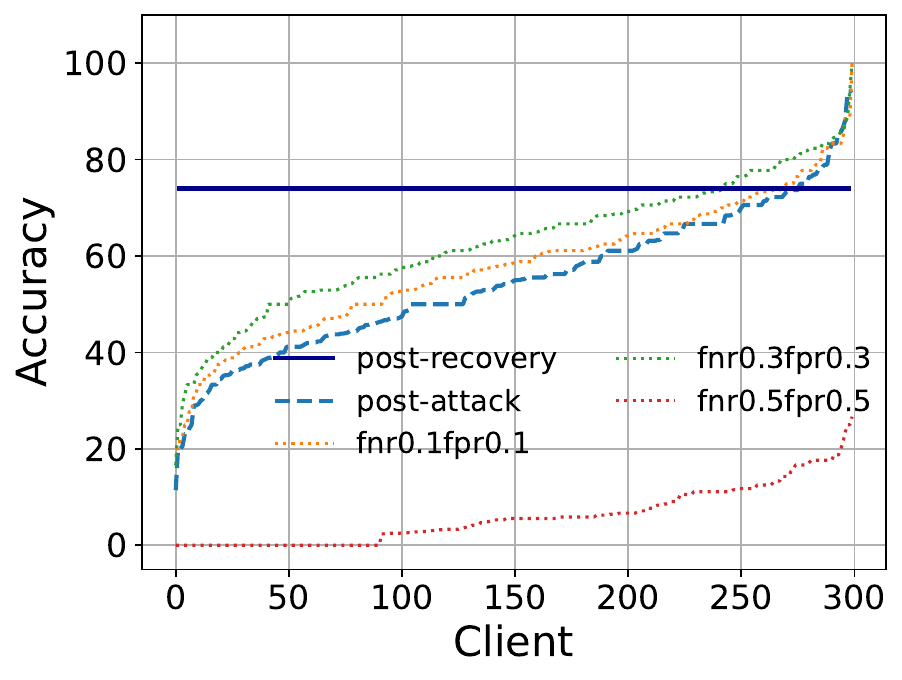}
        \caption{FedRecover}
        \label{fig:fedrecover_per_client_acc}
    \end{subfigure}
    \caption{Personalized evaluations, i.e., per-client accuracy for FEMNIST with TrMean and FedRecover. Note that the accuracy does not increase monotonically with the client number, rather we plot it in an ascending order here since order does not matter when we want to show variation in per-client accuracy. This should not be perceived as a relationship between accuracy and client ID/number. We are simply showing the range of individual client accuracies and have plotted them in a monotonic fashion for visual comparison.}
    \label{fig:per_client_acc}
\end{figure}

We train on FEMNIST for 200 rounds, achieving a global model accuracy of $82\%$. Figure~\ref{fig:trmean_per_client_acc} displays the accuracy trends, where \emph{global w/o attack} represents the global model accuracy, and \emph{per-client w/o attack} shows individual client test accuracies. \emph{Per-client w/o attack} clusters around the global accuracy, indicating the model learns from the combined data. Due to heterogeneity, we see a lot of variation in the performance of individual clients. In the attack scenario (\emph{global w/ attack} at $66\%$ and \emph{per-client w/ attack}), a similar trend persists, but most clients now fall below the global attack accuracy.
\emph{Based on our observations due to heterogeneity across clients, we strongly advocate using per-client metrics in future evaluations.}

\subsubsection{FedRecover's performance greatly varies on a per-client basis}\label{impact:evaluation:fdr}
Similar to the personalized evaluations for TrMean in \S\ref{impact:evaluation:trmean}, we apply personalized evaluations to FedRecover for the FEMNIST dataset. After performing FedRecover on the global model with $66\%$ accuracy (\S\ref{impact:evaluation:trmean}), we achieve a post-recovery accuracy of $74\%$ (\emph{post-recovery} in Figure~\ref{fig:fedrecover_per_client_acc}). This is illustrated on a per-client basis with the lines \emph{$fnr0.1fpr0.1$}, \emph{$fnr0.3fpr0.3$}, \emph{$fnr0.5fpr0.5$} in Figure~\ref{fig:fedrecover_per_client_acc}. The post-recovery accuracies for various FNRs and FPRs indicate that most client models fall below the global post-recovery accuracy. At an FNR and FPR of 0.5, all clients lie below both the post-recovery accuracy and the post-attack accuracy, signifying a failure in achieving recovery. This aligns with our findings in \S\ref{impact:attacks:fdr} that \emph{the presence of malicious clients during recovery leads to higher estimation errors, as mirrored in our per-client evaluations. This consistency underscores the necessity for personalized evaluations, particularly in non-i.i.d. datasets.}

\subsubsection{Impact of class-imbalance on per-class accuracies}\label{impact:evaluation:imbalance}

\begin{figure}
\centering
\includegraphics[scale=.5]{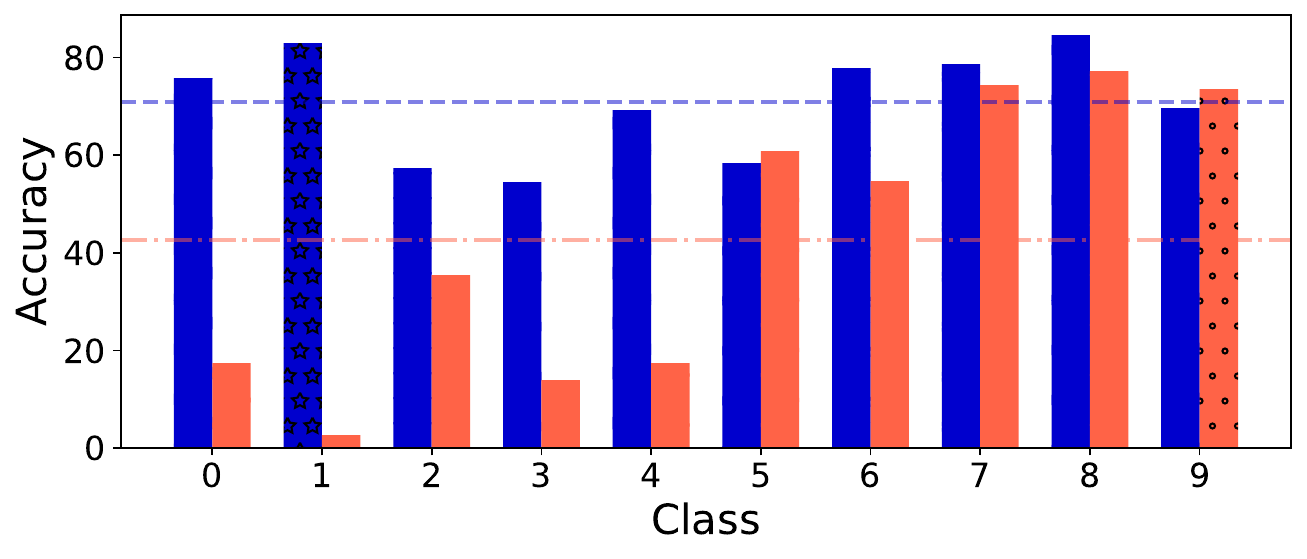}
\vspace*{-0.3cm}
\caption{Overall and per-class accuracies for balanced CIFAR10 before and after Stat-Opt attack.}
\label{fig:balanced_cifar}
\vspace*{-0.5cm}
\end{figure}

\begin{figure}
\centering
\includegraphics[scale=.5]{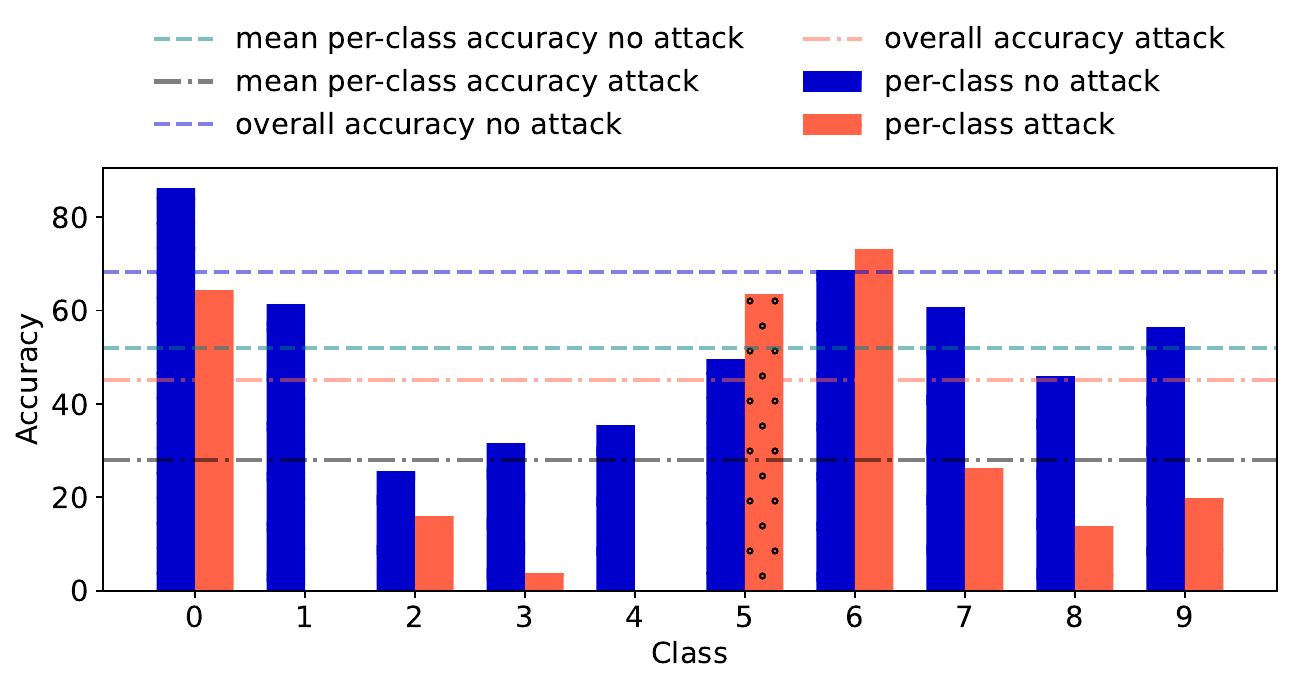}
\caption{Overall, per-class, and mean per-class accuracies for imbalanced CIFAR10 before and after Stat-Opt attack. Class 0 has the most samples.}
\label{fig:imbalanced_cifar}
\end{figure}

We examine the impact of class imbalance in CIFAR10 on the performance of TrMean under the no-attack and Stat-Opt attack scenarios. To set up our baseline, we train an Alexnet model with 100 clients over 100 epochs. We use the standard CIFAR10 dataset that has 50000 training samples and 10000 test samples.
Figure~\ref{fig:balanced_cifar} shows that we achieve an overall accuracy of $70.82\%$ without attack and $42.66\%$ with attack in the balanced setting. It also shows the respective per-class accuracies. We define the \emph{overall accuracy} as the total number of correct samples out of the total number of samples in the test dataset and the \emph{mean per-class accuracy} as the mean of all the per-class accuracies in the test dataset. In this case, since the test dataset is perfectly balanced, i.e., each class has the same number of test samples, the overall accuracy and the mean accuracy are the same.

Next, we create an imbalance in the CIFAR10 dataset by removing $90\%$ of the samples of all classes except class zero. This produces 5000 samples of class zero and 500 samples each for other classes in the training set.
Figure~\ref{fig:imbalanced_cifar} shows that in the imbalanced setting, the overall accuracy in no-attack is $68\%$, which is close to its balanced accuracy of $70.82\%$. However, this accuracy is biased towards class 0, which has the highest per-class accuracy of $86.2\%$($75.8\%$ in the balanced scenario) since it has ten times more samples than the rest. The mean per-class accuracy is $52\%$ as it ignores the class imbalance. This indicates that \emph{the mean per-class accuracy is a better metric than overall accuracy in this case, as it shows that the model loses utility with reduced samples.}

The imbalanced Stat-Opt scenario yields an overall accuracy of $48\%$, surpassing the balanced counterpart at $42.66\%$. Figure~\ref{fig:imbalanced_cifar} shows that the overall accuracy is significantly influenced by class 0, dropping from $86.2\%$ to $64.3\%$ under attack, compared to a drop from $75.8\%$ to $17.3\%$ in the balanced scenario. The mean per-class accuracy for the imbalanced attack is $27\%$. This highlights that \emph{dominant classes are less affected by attacks}, influencing the overall accuracy to reflect their performance. \emph{While overall accuracy is crucial, we stress the importance of reporting per-class and mean per-class accuracies, especially in real-world datasets with class imbalances, as neglecting this aspect can lead to misjudgments about a system's robustness.}
\section{Related Work}\label{sec:related}
\noindent\textbf{Previous systemizations: }
In prior research, various taxonomies
in adversarial ML have been developed, extending beyond
FL~\cite{goldblum2020dataset, jere2020taxonomy, barreno2010the, biggio2018wild, huang2011adversarial, rodriguez2023survey} and their main focus is on the attacks.
~\cite{rodriguez2023survey}  presents a taxonomy specifically for FL defenses, categorizing them based on their occurrence at the client, server, or communication channel. Our contribution extends beyond this by providing a more comprehensive and multidimensional systemization of the existing defenses in the literature. Additionally, we leverage our systemization to identify representative defenses for in-depth pitfall analyses.

Shejwalkar et al.~\cite{shejwalkar2022back} have performed the systemization of FL attacks, highlighting misconceptions about the robustness of FL systems that may arise from overlooking practical considerations about the threat model in deployment scenarios. Their conclusion emphasizes the high robustness of FL with simple and cost-effective defenses in practical threat models. Our work complements this perspective by focusing on defenses. We systematically categorize and re-evaluate representative defenses, uncovering common pitfalls, and providing actionable recommendations to address each.

\noindent\textbf{Pitfalls and guidelines: }
Arp et al.~\cite{arp2022and} have extensively studied the pitfalls associated with ML in computer security, identifying and providing recommendations for several challenges in this domain. It is crucial to emphasize the distinctions between our work and theirs. While we may identify similar pitfalls, such as Inappropriate Baseline (\S\ref{impact:algorithm}), Inappropriate Performance Measures (\S\ref{impact:evaluation}), and Lab-only Evaluation (\S\ref{impact:scalability}), the underlying reasons for these pitfalls and their impact are specific to FL. 
For example, a pitfall, ``inappropriate performance measures,'' addresses inappropriate performance measures like overall accuracy due to its limited capture of information about false positives and false negatives. In our context (\S\ref{impact:evaluation}), we discuss inappropriate performance measures, particularly in relation to fairness, and advocate for the use of personalized metrics, especially when dealing with non-i.i.d. data.

In the space of adversarial ML, Carlini et al.~\cite{carlini2019evaluating} identified numerous pitfalls associated with evaluating defenses against adversarial learning. While there are some commonalities with our work, such as considerations regarding the choice of attacks and testing against adaptive attacks, the primary distinction lies in the fact that our evaluation is specifically conducted in FL. Certain evaluation components are exclusive to FL, such as dealing with client heterogeneity and personalized evaluations.

\noindent\textbf{Improvements over~\cite{khan2023pitfalls}:} We take inspiration from~\cite{khan2023pitfalls}, which initiated the identification of FL pitfalls and used FedRecover as a case study to showcase the impact of some of these pitfalls. Before building upon their work, we set out to perform a systemization of FL defenses to understand the FL defenses space. This helps us to select representative defenses for pitfall analysis. We have expanded \cite{khan2023pitfalls}'s analysis by exploring additional pitfalls and studying their impact. Additionally, we broadened the evaluation spectrum by incorporating additional representative defenses like FLDetector and TrMean. Our study extends to a large-scale evaluation of FL with the language modality using StackOverflow~\cite{stackoverflow2019}.
\section{Limitations and Future Work}\label{sec:future}
\noindent\textbf{Scope of our evaluation: }
Performing a thorough exploration and evaluation of all the defenses in Table~\ref{tab:defenses}, along with additional ones, goes beyond the scope of a single paper. Performing a thorough exploration of one defense from each dimension, type, and attribute is also beyond the scope of one single paper as it requires running the full set of experiments for every defense and then comparing them.

\noindent\textbf{Expandability of our systemization: }
The defenses chosen for evaluation serve as representatives of the broader defense literature; however, it is acknowledged that alternative evaluations may differ based on different sets of defenses. Also, a newly designed defense based on a novel technique might not fit exactly along the attributes in our systemization. Therefore, our systemization remains adaptable (as detailed in \S\ref{sec:introduction}), allowing for the incorporation of additional type, attributes, and even dimensions in the future.

\noindent\textbf{Incorporating other modalities:}
As highlighted in \S\ref{sec:introduction}, our aim is to expand the scope of our evaluation across diverse dimensions, encompassing data distributions, data modalities, and the nature of the ML task. While we have addressed image classification and NLP, other modalities, such as multimodal time-series tasks~\cite{zhao2022multimodal}, and the emerging paradigm of \emph{vision-language models}~\cite{radford2021learning}, remain unexplored. The evaluation of these modalities, along with more complex vision-language models, under traditional threat models could prove intriguing. Such exploration might lead to the development of new attacks and defenses, potentially uncovering novel pitfalls in the process.

\section{Conclusion}\label{sec:conclusion}
Our study contributes a crucial systemization of FL defenses, offering valuable insights for researchers and practitioners in the selection, combination, and design of defenses.   
Additionally, we address an often-overlooked aspect in the FL poisoning defense literature— the \emph{experimental setups employed} to assess defense efficacy.  
After reviewing a variety of defense works, we identify prevalent questionable experimental trends. Through case studies featuring well-known defenses such as TrMean, FLDetector, and FedRecover, we illustrate how the choice of experimental setups can significantly impact robustness claims.

\newpage

\bibliographystyle{ACM-Reference-Format}
\bibliography{privacy}

\appendix
\section{Appendix}\label{sec:appendix}

\subsection{Our methodology to classify 50 defenses}\label{appdx:classification_method}
Table~\ref{tab:defenses} shows our survey of the 50 defense works and our classification methodology along the six dimensions: Datasets, Attacks, Data Distribution, FL-Algorithm, FL Type, and Evaluation Type.
\begin{table*}[]
\caption{Classification of 50 defense works across 6 dimensions of evaluation setup.}
\label{tab:defenses}
\scriptsize
\begin{tabular}{|l|p{2cm}|p{2.7cm}|l|p{1.5cm}|p{.7cm}|l|}
\hline
\textbf{Work}                                   & \textbf{Datasets}    & \textbf{Attacks}                 & \textbf{Data Distribution} & \textbf{FL Algorithm} & \textbf{FL Type} & \textbf{Evaluation} \\ \hline
FLDetector~\cite{zhang2022fldetector}                                      & FA,FE,C10            & Stat-Opt                             & Fang, Natural              & FedSGD                & CS               & Global              \\ \hline
FedRecover   ~\cite{cao2022fedrecover}                                   & M,FA,P,H             & Stat-Opt                             & Fang                       & FedSGD                & CS               & Global              \\ \hline
Machine Learning with Adversaries~\cite{blanchard2017machine}               & M, spambase          & RGA                              & IID                        & FedSGD                & CS               & Global              \\ \hline
FLTrust~\cite{cao2020fltrust}                                         & M,CHM,C10,H          & Krum, Stat-Opt, LF                   & Fang                       & FedAvg                & CS               & Global              \\ \hline
Byzantine-Robust Distributed Learning~\cite{yin2018byzantine}           & M                    & RGA                              & IID                        & FedSGD                & CS               & Global              \\ \hline
Provably Secure Federated Learning against~\cite{cao2021provably}      & M                    & Not applicable                   & Fang                       & FedAvg                &                  & Global              \\ \hline
Learning to Detect Malicious Clients for~\cite{li2020learning}        & M, FE, S140          & SF, AN, BD                       & Natural, McMahan           & FedAvg                & CD               & Global              \\ \hline
Robust Federated Learning~\cite{xie2022robust}                       & M,FE,C10/100,N20     & LF, LIE, Fang                    & Dirichlet, Natural         & FedAvg                & CD               & Global              \\ \hline
The Hidden Vulnerability of Distributed~\cite{mhamdi2018the}         & M, C10               & Specific attack                  & IID                        & FedSGD                & CS/CD            & Global              \\ \hline
Sageflow~\cite{park2021sageflow}                                        & M, FA, C10           & SF, LF                           & McMahan                    & FedAvg                & CS               & Global              \\ \hline
Mitigating Irrelevant Clients in FL~\cite{nagalapatti2021game}             & M                    & LF                               & McMahan                    & FedAvg                & CS               & Global              \\ \hline
Cronus~\cite{chang2019cronus}                                          & M, C10, P, Svhn      & \{LF, LIE\}                      & IID                        & FedAvg                & CS               & Global              \\ \hline
Can You Really Backdoor Federated Learning?~\cite{sun2019can}     & FE                   & BD                               & Natural                    & FedAvg                & CS/CD            & Global              \\ \hline
Generalized Byzantine-tolerant SGD~\cite{xie2018generalized}              & M, C10               & BF, LF, LIE                      & IID                        & FedSGD                & CS               & Global              \\ \hline
The Limitations of Federated Learning~\cite{fung2020limitations}           & M, VGG, KDD, A       & LF, BD                           & Each class to a client     & Both                  & CS               & Global              \\ \hline
Auror~\cite{shen2016auror}                                           & M                    & Targeted-LF                      & IID                        & FedSGD                & CS               & Global              \\ \hline
Robust Aggregation for Federated Learning~\cite{pillutla2019robust}       & FE,S140,S            & Specific attacks, RGA            & Natural                    & FedAvg                & CS/CD            & Global              \\ \hline
CRFL~\cite{xie2021crfl}                                            & M, FE                & BD                               & IID                        & FedAvg                & CS               & Global              \\ \hline
FLIP~\cite{zhang2022flip}                                            & M, FA, C10           & BD                               & Dirichlet                  & FedAvg                & CS               & Global              \\ \hline
RoFL~\cite{burkhalter2021rofl}                                            & FE, C10              & BD                               & Natural, Dirichlet         & FedAvg                & CD               & Global              \\ \hline
Securing FL against Overwhelming~\cite{ranjan2022securing}                & M, FA                & LF, BD                           & Dirichlet                  & FedAvg                & CS               & Global              \\ \hline
Defending against the Label-flipping Attack~\cite{jebreel2022defending}     & M, C10               & LF,                              & IID, Dirichlet             & FedAvg                & CS               & Global              \\ \hline
FRL~\cite{mozaffari2021frl}                                             & M, FE, C10           & Fang, Dyn-Opt                     & Dirichlet, Natural         & FedAvg                & CD               & Global              \\ \hline
CONTRA~\cite{awan2021contra}                                          & M, C10, Loan         & LF, BD                           & Dirichlet                  & FedAvg                & CS               & Global              \\ \hline
EIFFeL~\cite{roy2022eiffel}                                          & M, FA, FE,C10        & LIE, RGA, SF, Dyn-Opt             & IID, Natural               & FedAvg                & CS/CD            & Global              \\ \hline
Local and central DP for robustness~\cite{naseri2020local}         & E, C10, s140, Reddit & BD                               & McMahan                    & FedAvg                & CS/CD            & Global              \\ \hline
Signguard~\cite{xu2021signguard}                                       & M, FA, C10, AGnews   & LIE, RGA, SF, Dyn-Opt             & IID                        & FedSGD                & CS               & Global              \\ \hline
DisBezant~\cite{ma2022disbezant}                                       & \{M, FA, C10\}       & RGA                              & \{Fang\}                   & FedAvg                &                  & Global              \\ \hline
Learning from History for Byzantine~\cite{praneeth2020learning}             & M, C10               & BF, LF, LIE                      & Exponential                & FedSGD                & CS               & Global              \\ \hline
Byzantine-robust learning on heterogeneous~\cite{karimireddy2020byzantine}      & M                    & BF, LF, LIE, IPM, Mimic          & \{McMahan\}                & FedSGD                & CS               & Global              \\ \hline
Byzantine-Resilient Non-Convex Stochastic~\cite{allen2020byzantine}       & C10, C100            & SF, LF, LIE, Delayed-grad        & IID                        & FedSGD                & CS               & Global              \\ \hline
Byzantine-robust Federated Learning~\cite{li2021byzantine}          & M, FA, C10,Spambase  & LF, IPM, LIE, Uniform, arbitrary & McMahan                    & FedAvg                & CS/CD            & Global              \\ \hline
Stochastic alternating direction method of~\cite{lin2021stochastic}      & M, Covertype         & RGA, SF, LF                      & IID                        & FedSGD                & CS               & Global              \\ \hline
Variance reduction is an antidote to byzantines~\cite{gorbunov2022variance} & LIBSVM               & LF, BF, LIE, IPM                 & \{IID\}                    & FedSGD                & CS               & Global              \\ \hline
On the byzantine robustness of clustered FL~\cite{sattler2020byzantine}     & M, FA, C10           & RGA, LF, Uniform noise           & IID                        & FedSGD                & CS               & Global              \\ \hline
RSA~\cite{li2019rsa}      & M                    & SF                               & IID                        & FedSGD                & CS               & Global              \\ \hline
Federated variance-reduced~\cite{wu2020federated}                      & ijcnn1, covtype      & RGA, SF, Zero-grad               & IID                        & FedSGD                & CS               & Global              \\ \hline
Abnormal client behavior detection in~\cite{li2019abnormal}           & FE                   & SF, RGA, Grad ascent             & Natural                    & FedAvg                & CD               & Global              \\ \hline
Distributed Momentum for Byzantine~\cite{el2021distributed}              & M, FA, C10/100       & LIE, IPM                         & IID                        & FedSGD                & CS               & Global              \\ \hline
Attack-resistant FL with residual-based~\cite{fu2019attack}         & M, C10, Amazon, Loan & LF, BD                           & Dirichlet, Natural         & FedAvg                & CS               & Global              \\ \hline
Towards communication-efficient~\cite{liu2021towards}                 & M                    & LF                               & IID                        & FedSGD                & CS               & Global              \\ \hline
Justinian's GAAvernor~\cite{pan2020justinian}                           & M, C10, Yelp, Health & RGA                              & IID                        & FedSGD                & CS               & Global              \\ \hline
Untargeted poisoning attack detection~\cite{mallah2021untargeted}           & M, C10, MTL Trajet   & BD                               & IID                        & FedSGD                &                  & Global              \\ \hline
TDFL~\cite{xu2022tdfl}                                            & M, FA, C10           & LF, RGA, Krum, Stat-Opt, BD          & McMahan                    & FedAvg                & CD               & Global              \\ \hline
Siren~\cite{guo2021siren}                                           & FA, C10              & SF, LF, bhagoji                  & Fang                       & FedAvg                & CS               & Global              \\ \hline
FLARE~\cite{wang2022flare}                                           & FA, C10, Kather      & Krum, Stat-Opt                       & IID                        & FedAvg                & CS               & Global              \\ \hline
Analyzing Federated Learning Through~\cite{bhagoji2019analyzing}            & FA, UCI Census       & Specific attack                  & IID                        & FedAvg                &                  & Global              \\ \hline
BaFFLe~\cite{andreina2021baffle}                                          & C10, FE              & BD                               & Dirichlet                  & FedAvg                &                  & Global              \\ \hline
Defending against backdoors in FL~\cite{ozdayi2021defending}               & FA, FE               & BD                               & Both                       & FedAvg                &                  & Global              \\ \hline
Ditto~\cite{li2021ditto}                                           & FA, FE, CelebA       & LF, RGA, BD                      & Natural, McMahan           & FedAvg                & CS               & Personalized        \\ \hline
\end{tabular}
\end{table*}

\subsection{Full names for datasets and attacks}\label{appdx:abbreviations}

\begin{table}[t]
\centering
\scriptsize
\caption{Abbreviations and full-forms of datasets in Table~\ref{tab:defenses}.}
\label{tab:full_names_datasets}
\begin{tabular}{|l|l|}
\hline
M    & MNIST                     \\ \hline
FA   & FashionMNIST              \\ \hline
FE   & FEMNIST                   \\ \hline
C10  & CIFAR10                   \\ \hline
C100 & CIFAR100                  \\ \hline
P    & Purchase                  \\ \hline
H    & HAR                       \\ \hline
S140 & Sentiment140              \\ \hline
SVHN & Street-view House Numbers \\ \hline
VGG  & VGGFace                   \\ \hline
KDD  & KDDCup                    \\ \hline
N20  & News 20                   \\ \hline
A    & Amazon                    \\ \hline
S    & Shakespeare               \\ \hline
\end{tabular}
\end{table}

\begin{table}[H]
\centering
\scriptsize
\caption{Abbreviations and full-forms of attacks in Table~\ref{tab:defenses}.}
\label{tab:full_names_attacks}
\begin{tabular}{|l|l|}
\hline
LF  & Label Flip                 \\ \hline
IPM & Inner Product Manipulation \\ \hline
SF  & Sign Flip                  \\ \hline
BF  & Bit Flip                   \\ \hline
LIE & Little is Enough           \\ \hline
AN  & Additive Noise             \\ \hline
BD  & Backdoor~\cite{bagdasaryan2018how} \\ \hline
RGA & Random Gaussian Attack \\ \hline
\end{tabular}
\end{table}
\end{document}